\documentclass[aps,prd,twocolumn,preprintnumbers,superscriptaddress,nofootinbib, amsmath,amssymb]{revtex4-1}

\usepackage{graphicx,subfigure}
\usepackage{dcolumn}
\usepackage{bm}
\usepackage{mathrsfs}
\usepackage{enumitem}
\usepackage{times}
\usepackage{hyperref}
\hypersetup{
    colorlinks=true,
    linkcolor=blue,
    citecolor=magenta,
    filecolor=magenta,
    urlcolor=blue
}

\def\beq{\begin{equation}} \def\eeq{\end{equation}}
\def\bear{\begin{eqnarray}} \def\ear{\end{eqnarray}}

\begin{document}

\title{Exploring black hole shadows in axisymmetric spacetimes with coordinate-independent methods and neural networks}

\author{Temurbek~Mirzaev}
\email{mtemur141096@gmail.com} 
\affiliation{Center for Astronomy and Astrophysics, Center for Field Theory and Particle Physics, and Department of Physics,\\
Fudan University, Shanghai 200438, China}
\affiliation{Institute for Advanced Studies, New Uzbekistan University, Movarounnahr str. 1, Tashkent 100007, Uzbekistan}

\author{Bobomurat Ahmedov}
\email{ahmedov@astrin.uz} 
\affiliation{School of Physics, Harbin Institute of Technology, Harbin 150001, People’s Republic of China}
\affiliation{Institute for Advanced Studies, New Uzbekistan University, Movarounnahr str. 1, Tashkent 100007, Uzbekistan}
\affiliation{Institute of Theoretical Physics, National University of Uzbekistan, Tashkent 100174, Uzbekistan}

\author{Cosimo~Bambi}
\email{bambi@fudan.edu.cn} 
\affiliation{Center for Astronomy and Astrophysics, Center for Field Theory and Particle Physics, and Department of Physics,\\
Fudan University, Shanghai 200438, China}
\affiliation{School of Natural Sciences and Humanities, New Uzbekistan University, Tashkent 100007, Uzbekistan}

\begin{abstract}
The study of black hole shadows provides a powerful tool for testing the predictions of general relativity and exploring deviations from the standard Kerr metric in the strong gravitational field regime. Here, we investigate the shadow properties of axisymmetric gravitational compact objects using a coordinate-independent formalism. We analyze black hole shadows in various spacetime geometries, including the Kerr, Taub-NUT, $\gamma$, and Kaluza-Klein metrics, to identify distinctive features that can be used to constrain black hole parameters. To achieve a more robust characterization, we employ both Legendre and Fourier expansions, demonstrating that the Fourier approach may offer better coordinate independence and facilitate cross-model comparisons. Finally, we develop a machine learning framework based on neural networks trained on synthetic shadow data, enabling precise parameter estimation from observational results. Using data from observational astronomical facilities such as the Event Horizon Telescope (EHT), Keck, and the Very Large Telescope Interferometer (VLTI), we provide constraints on black hole parameters derived from shadow observations. Our findings highlight the potential of coordinate-independent techniques and machine learning for advancing black hole astrophysics and testing fundamental physics beyond general relativity.
\end{abstract}

\maketitle

\section{Introduction}

 The study of black hole shadows has emerged as a key observational tool for testing the foundations of general relativity (GR) and exploring the nature of spacetime in extreme gravitational environments~\cite{EHT2019_M87I,EHT_SgrA_I, 2010ApJ...716..187J, BambiFreese2009, Wielgus2020,Perlick2022}. The gravitational field surrounding black holes is so strong that even photons can orbit in unstable trajectories, forming what is known as the photon sphere~\cite{Claudel2001}. The black hole shadow, an observable manifestation of the photon sphere, offers a unique opportunity to explore the geometry of spacetime around black holes.
The first direct imaging of a black hole shadow, based on 2017 observations and published by the Event Horizon Telescope (EHT) collaboration in 2019, revealed the shadow of the supermassive black hole M87*, marking a pivotal breakthrough in astrophysics~\cite{EHT_Fi}.
Although the observations of Sgr A* were conducted in 2017, the EHT collaboration published the results more recently, in 2022, expanding our capabilities  to test predictions of GR  ~\cite{EHT_SgrA_I} 
and alternative gravitational theories, as well as to examine the very nature of black hole spacetimes in the strong-field regime \cite{tests1, tests2, tests3, EHT_SgrA_VI}.
These developments have spurred the search for new ways to characterize black hole shadows and constrain fundamental parameters of black holes.

Characterizing the morphology of black hole shadows accurately is crucial, not only for verifying GR but also for probing deviations indicative of physics beyond the standard Kerr paradigm~\cite{BambiFreese2009,Johannsen2010,Konoplya2016,CunhaHerdeiro2018,Kocherlakota2021}. Indeed, numerous alternative spacetime geometries, including charged, scalarized, or higher-dimensional solutions, have been proposed, each predicting subtle yet distinctive differences in shadow size and shape~\cite{Kocherlakota_2021, Tsukamoto_2024}. Constraining these deviations observationally could provide vital clues to potential extensions of fundamental theories, thus motivating the exploration of various non-Kerr metrics such as Johannsen-Psaltis, Konoplya-Rezzolla-Zhidenko, $\gamma$, Taub-NUT, and Kaluza-Klein spacetimes alongside the Kerr solution, which serves as the cornerstone model for astrophysical black holes~\cite{kerr-hypothesis1, kerr-hypothesis2, kerr-hypothesis3, Tao_2023,Abdujabbarov13c, kaluzaklein}.

Despite substantial progress, characterizing black hole shadows across different metrics has typically relied on specific coordinates to represent shadow shapes, making it challenging to directly compare theoretical predictions with observational data from distant observers.
This challenge motivates the development of unified, coordinate-independent methods. A notable step in this direction was made by Abdujabbarov et al.~\cite{Abdujabbarov_15}, who introduced a formalism based on Legendre polynomial expansions to describe the black hole shadow in a coordinate-free manner.  While the Legendre expansion effectively represents shadows, it lacks robustness under rotational transformations. In this work, we build upon this idea and introduce an alternative approach based on Fourier series, which offers intrinsic invariance under rotations and is therefore well suited for comparing shadow features across a wider range of spacetime geometries.

Motivated by these considerations, we adopt and extend the coordinate-independent framework, emphasizing the Fourier approach, to characterize black hole shadows comprehensively. We systematically examine shadow features across different axisymmetric gravitational metrics, such as Kerr~\cite{met-kerr}, $\gamma$~\cite{met-gam1, met-gam2, met-gam3}, rotating $\gamma$ ~\cite{met-delta1, met-delta2}, Taub-NUT~\cite{met-taub1, met-taub2}, and Kaluza-Klein~\cite{met-kal_th, met-kalm} spacetimes. Our investigation employs both numerical ray-tracing methods and analytical results (where available) to establish robust and precise benchmarks. The results are subsequently integrated with current observational constraints obtained from advanced astronomical facilities, including Keck, the Very Large Telescope Interferometer (VLTI), and the EHT~\cite{Keck_2019, VLTI_2021, VLTI_2022, EHT_Fi}.

Furthermore, we introduce a machine learning methodology based on neural networks to facilitate rapid, accurate, and model-independent extraction of black hole parameters directly from shadow observations. By training neural networks on extensive synthetic datasets generated through careful numerical simulations across a broad parameter space we develop a tool capable of efficiently interpreting observational data. The neural network learns the intricate relationship between coordinate-independent shadow descriptors and underlying spacetime parameters, significantly streamlining the inference process and enabling precise constraints on potential deviations from GR.

This paper is structured as follows. In Section~\ref{sec:shadows}, we provide a detailed theoretical overview of black hole shadows, focusing on numerical ray-tracing algorithms and analytic treatments within various axisymmetric spacetime metrics. Section~\ref{sec:char} thoroughly introduces the coordinate-independent characterization of black hole shadows, contrasting the Legendre and Fourier expansions, and emphasizes the utility of the rotation-invariant Fourier approach. Section~\ref{sec:nn} presents our neural network-based framework, demonstrating its effectiveness in extracting physical parameters from shadow data and illustrating its reliability through extensive validation tests. Finally, Section~\ref{sec:conclusion} summarizes our key results and discusses future avenues for extending these techniques to other gravitational theories and astrophysical scenarios.
For completeness, the line elements of the spacetime metrics used throughout the paper are listed in Appendix~\ref{metric}.

\section{Shadow of the Black Hole for Different Metrics and Constraints Using Observational Data} \label{sec:shadows}

\subsection{Ray Tracing}

This section outlines a numerical algorithm used to trace photon trajectories from an observer’s image plane at infinity to a black hole, following methods established in previous studies~\cite{Psaltis_2011, 2010ApJ...716..187J, kaluzaklein, PhysRevD.106.084041}. Adopting the approach of~\cite{2007ApJ...654..458C}, the trajectory's time and azimuthal coordinates are defined through first-order differential equations involving two conserved quantities, while the radial and polar geodesic motion is described by second-order differential equations. The photon sphere in a black hole spacetime acts as a boundary, separating geodesics that escape to infinity from those falling into the event horizon~\cite{Claudel2000TheGO}. The black hole shadow corresponds to the observed projection of that photon sphere. Since null-geodesic equations cannot generally be solved analytically in arbitrary spacetimes, numerical methods are required to compute the photon motion and determine the shadow. 
However, in Kerr spacetime, the photon sphere can be described analytically, allowing for a direct calculation of the shadow boundary from the known photon orbit solutions.


To derive the equations for the $t$- and $\phi$-components of the photon position, 
we adopt a Boyer–Lindquist–like coordinate system $(t,r,\theta,\phi)$ with metric 
signature $(-,+,+,+)$. In this convention, stationary and axisymmetric (circular) 
spacetimes admit two Killing vectors associated with time-translation and axial 
symmetry, respectively:
$\xi^\mu = (1,0,0,0)$ and $\eta^\mu = (0,0,0,1)$. These symmetries lead to the 
conservation of energy and angular momentum along the photon trajectory, expressed as:

\begin{equation}
E = -g_{tt} \frac{d t}{d \lambda} - g_{t \phi} \frac{d \phi}{d \lambda},
\end{equation}

\begin{equation}
L_z = g_{\phi \phi} \frac{d \phi}{d \lambda} + g_{t \phi} \frac{d t}{d \lambda},
\end{equation}

where $g_{\mu \nu}$ denotes the metric components and $\lambda$ is an affine parameter. These conserved quantities allow for the derivation of two first-order differential equations describing the evolution of the $t$ and $\phi$ components of the photon position:

\begin{equation}
\frac{d t}{d \lambda^{\prime}} = \frac{-g_{\phi \phi} - b_o g_{t \phi}}{g_{\phi \phi} g_{t t} - g_{t \phi}^2},
\end{equation}

\begin{equation}
\frac{d \phi}{d \lambda^{\prime}} = \frac{b_o g_{t t} + g_{t \phi}}{g_{\phi \phi} g_{t t} - g_{t \phi}^2},
\end{equation}

where the normalized affine parameter is defined as $\lambda' \equiv E \lambda$, and the impact parameter of the photon trajectory is given by $b_o \equiv L_z / E$.

The $r$- and $\theta$-components of the photon position are obtained by solving the second-order geodesic equations for a generic circular metric:

\begin{equation}\label{geod1}
	\begin{aligned}
		& \frac{d^2r}{d\lambda'^2}=
		-\Gamma_{tt}^{r}\left(\frac{dt}{d\lambda'}\right)^2
		-\Gamma_{rr}^{r}\left(\frac{dr}{d\lambda'}\right)^2
		-\Gamma_{\theta\theta}^{r}\left(\frac{d\theta}{d\lambda'}\right)^2 \\
		&  \ \ \ \ \ \ \ \ \ \ -\Gamma_{\phi\phi}^{r}\left(\frac{d\phi}{d\lambda'}\right)^2
		-2\Gamma_{t\phi}^{r}\left(\frac{dt}{d\lambda'}\right)\left(\frac{d\phi}{d			\lambda'}\right) \\
		& \ \ \ \ \ \ \ \ \ \ -2\Gamma_{r\theta}^{r}\left(\frac{dr}{d\lambda'}\right)			\left(\frac{d	\theta}{d\lambda'}\right), \\
			\end{aligned}
\end{equation}
\begin{equation}\label{geod2}
	\begin{aligned}
		& \frac{d^2\theta}{d\lambda'^2}=
		-\Gamma_{tt}^{\theta}\left(\frac{dt}{d\lambda'}\right)^2
		-\Gamma_{rr}^{\theta}\left(\frac{dr}{d\lambda'}\right)^2 - 			\Gamma_{\theta\theta}^{\theta}\left(\frac{d\theta}{d\lambda'}\right)^2 \\
		&  \ \ \ \ \ \ \ \ \ \ -\Gamma_{\phi\phi}^{\theta}\left(\frac{d\phi}{d\lambda'}\right)^2 
		-2\Gamma_{t\phi}^{\theta}\left(\frac{dt}{d\lambda'}\right)\left(\frac{d			\phi}{d\lambda'}\right) \\
		& \ \ \ \ \ \ \ \ \ \  -2\Gamma_{r\theta}^{\theta}\left(\frac{dr}{d				\lambda'}\right)	\left(\frac{d\theta}{d\lambda'}\right),
	\end{aligned}
\end{equation}
%


Each photon is traced backward in time from the observer’s image plane (or “screen”), 
which we place at a large distance $d \gg M$ from the black hole (see Fig.~\ref{fig:screen_setup}). At such distances 
the spacetime curvature is negligible and the metric reduces to flat space, allowing 
the screen to be treated as a Cartesian plane. The screen is tilted by the inclination 
angle $i$ with respect to the black hole’s spin axis.
We parametrize the screen using celestial coordinates $(\alpha,\beta)$, where 
$\alpha$ measures the displacement along the horizontal axis and $\beta$ along the vertical axis. 
These screen coordinates are directly related to the photon’s impact parameters 
and enter the initial conditions of the geodesics through Eqs.~(\ref{in_pos_r})–(\ref{in_momen_t}). 
The initial position and four-momentum in Boyer–Lindquist coordinates are then 
computed as~\cite{Bambi17e}:

\begin{figure}[t]
\centering
\includegraphics[width=0.4\textwidth]{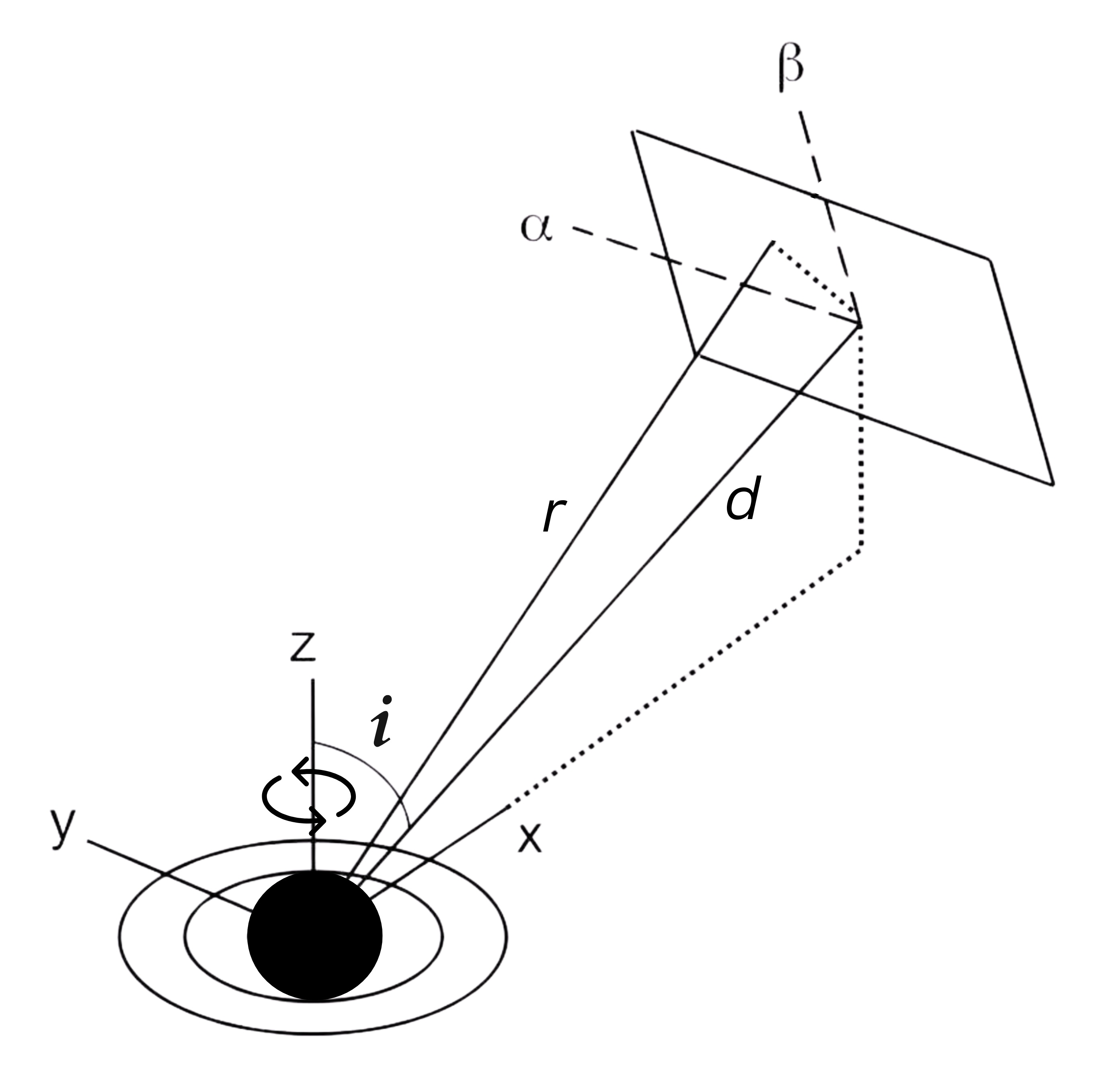}
\caption{
Ray-tracing setup: photons are launched from the observer’s screen at distance $d$, 
inclined by angle $i$ relative to the spin axis. The screen is parametrized by coordinates 
$(\alpha,\beta)$, which map to the photon impact parameters used in the geodesic integration.}
\label{fig:screen_setup}
\end{figure}

\begin{equation}\label{in_pos_r}
	r_j =\left(d^2+\alpha^2+{\beta}^2\right)^{1/2},
\end{equation}

\begin{equation}\label{in_pos_theta}
	\theta_j = \arccos\left(\frac{d\cos{i}+{\beta}\sin{i}}{r_j}\right),
\end{equation}

\begin{equation}\label{in_pos_phi}
	\phi_j = \arctan\left(\frac{{\alpha}}{d\sin{i}-{\beta}\cos{i}}\right),
\end{equation}

\begin{equation}\label{in_momen_r}
	\left(\frac{dr}{d\lambda'}\right)_j=\frac{d}{r_j},
\end{equation}

\begin{equation}\label{in_momen_theta}
	\left(\frac{d\theta}{d\lambda'}\right)_j = \frac{-\cos{i}+\frac{d}{r_j^2}(d\cos{i}+{\beta}\sin{i})}{\sqrt{r_j^2-(d\cos{i}+{\beta}\sin{i})^2}},
\end{equation}

\begin{equation}\label{in_momen_phi}
	\left(\frac{d\phi}{d\lambda'}\right)_j = \frac{-{\alpha}\sin{i}}{{\alpha}^2+(d\sin{i}-{\beta}\cos{i})^2},
\end{equation}


\begin{equation}\label{in_momen_t}
\scalebox{0.80}{$\displaystyle
\begin{split}
&\left(\frac{dt}{d\lambda'}\right)_j
= -\frac{g_{t\phi}}{g_{tt}}
   \left(\frac{d\phi}{d\lambda'}\right)_j \\
&\quad + \sqrt{
    \frac{g_{t\phi}^2}{g_{tt}^2}
    \left(\frac{d\phi}{d\lambda'}\right)_j^2
    - \left[
        \frac{g_{rr}}{g_{tt}}
        \left(\frac{dr}{d\lambda'}\right)_j^2
        + \frac{g_{\theta\theta}}{g_{tt}}
        \left(\frac{d\theta}{d\lambda'}\right)_j^2
        + \frac{g_{\phi\phi}}{g_{tt}}
        \left(\frac{d\phi}{d\lambda'}\right)_j^2
      \right]
},
\end{split}
$}
\end{equation}

Once all photon paths are traced from the observer’s screen, we identify the shadow by selecting those rays that orbit the black hole the most before falling in. These photons orbit many times in the azimuthal direction and in the polar angle. The boundary between escaping and captured photons, traced across all directions, then gives the closed shadow contour.

\subsection{Black Hole Shadow in Different Metrics and Observational Constraints}

Using ray tracing techniques explained in the previous section, we calculate the black hole shadow in axisymmetric spacetimes numerically. We consider different spacetimes: Kerr, $\gamma$, Taub-NUT, and Kaluza-Klein. The line elements for these spacetimes are provided in Appendix~\ref{metric}. For the Kerr and Taub-NUT metrics, the shadow can be obtained analytically. Since the shadow error can be controlled through ray tracing (with a target error in the shadow contour set to $\texttt{tol} = 1.0 \times 10^{-6}$), analytical solutions also help verify the precision of the numerical implementation.

In each case, we generate a large number of shadows by varying both the black hole parameters and the observer’s inclination angle, typically more than a million shadows per metric. While shadow generation in Kerr and Taub-NUT spacetime is computationally fast due to analytic solutions, the process is significantly slower for the $\gamma$ and Kaluza-Klein cases, where full numerical ray tracing is required.

\begin{figure*}[ht!]
    \centering
    \includegraphics[width=0.49\linewidth]{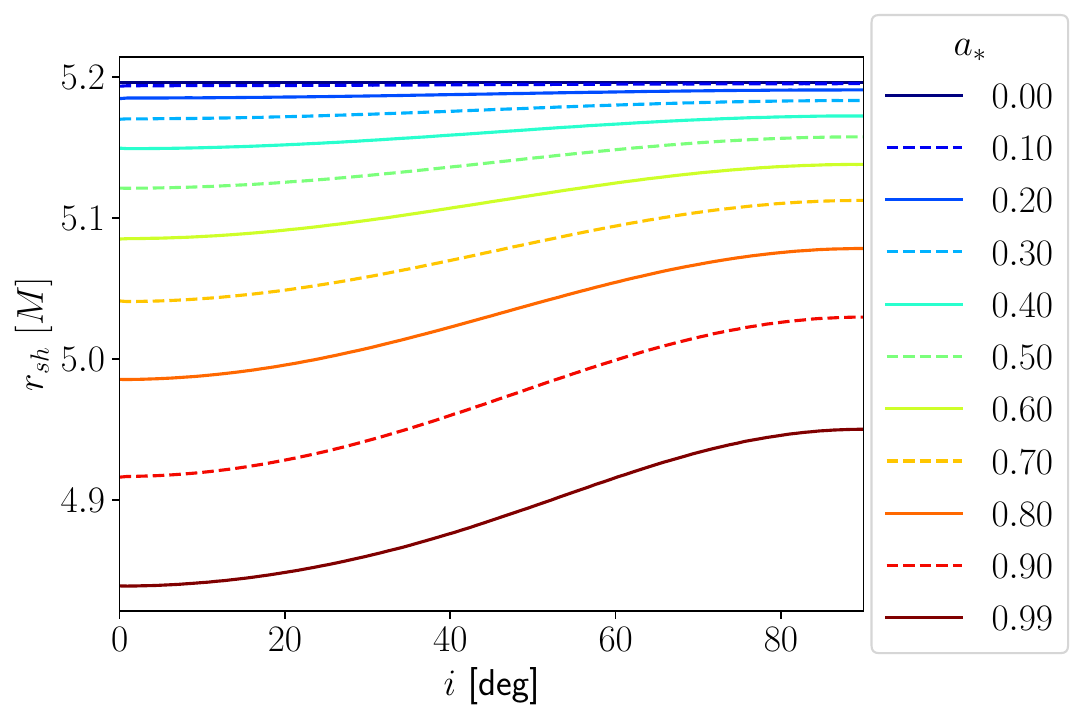}
    \includegraphics[width=0.49\linewidth]{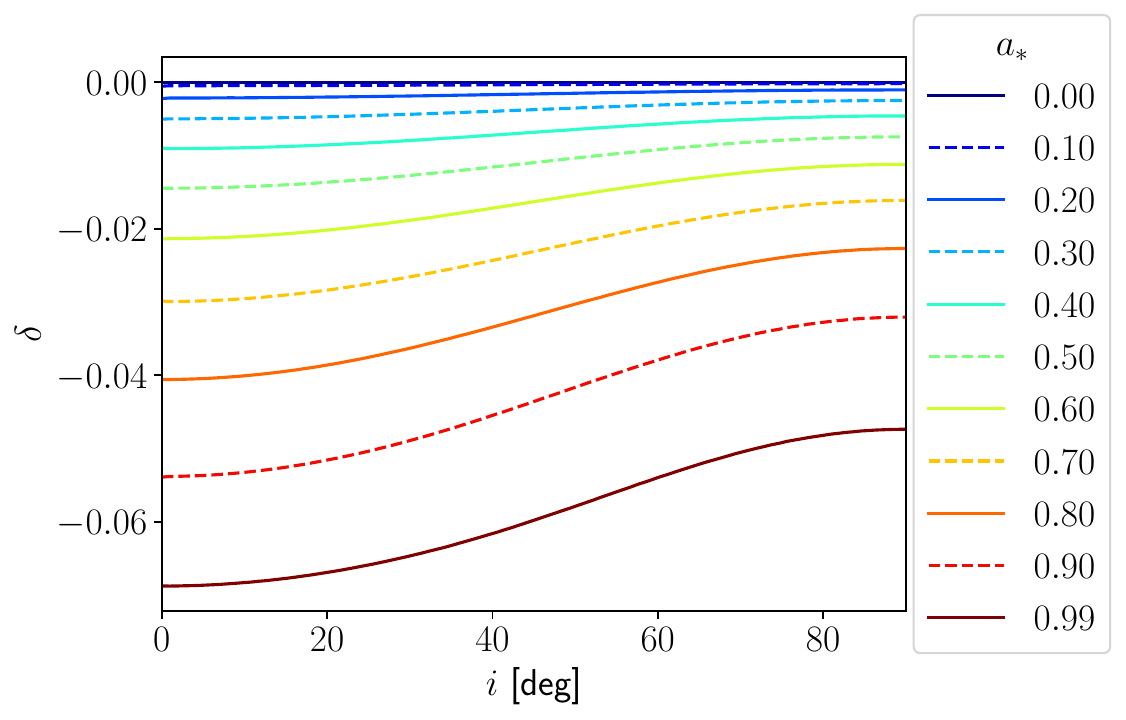}
    \caption{Shadow areal radius $r_{\rm sh}$ (left) and Schwarzschild deviation parameter $\delta$ (right) of a Kerr black hole as functions of observer inclination angle $i$ for spin parameters $a_* = 0.00,\,0.10,\,0.20,\,\dots,\,0.99$. Curves are colour-coded from blue ($a_*=0$) to dark red ($a_*=0.99$) and use alternating solid/dashed styles for clarity.}
    \label{kerr_sh}
\end{figure*}

\begin{figure*}[ht!]
    \centering
    \includegraphics[width=0.49\linewidth]{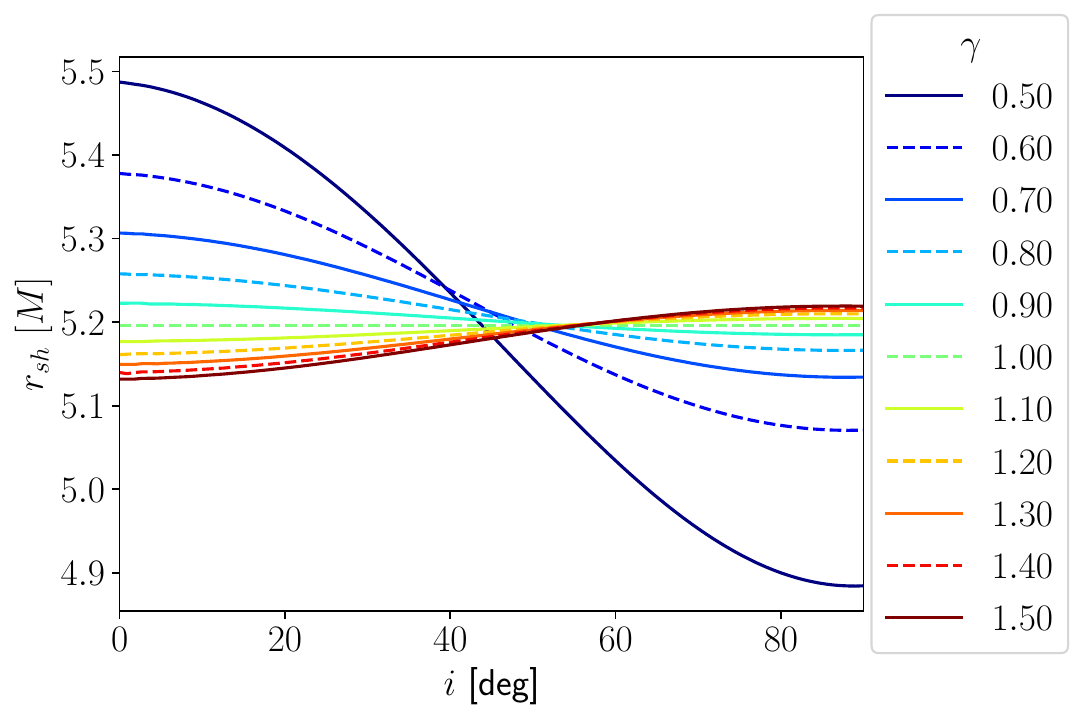}
    \includegraphics[width=0.49\linewidth]{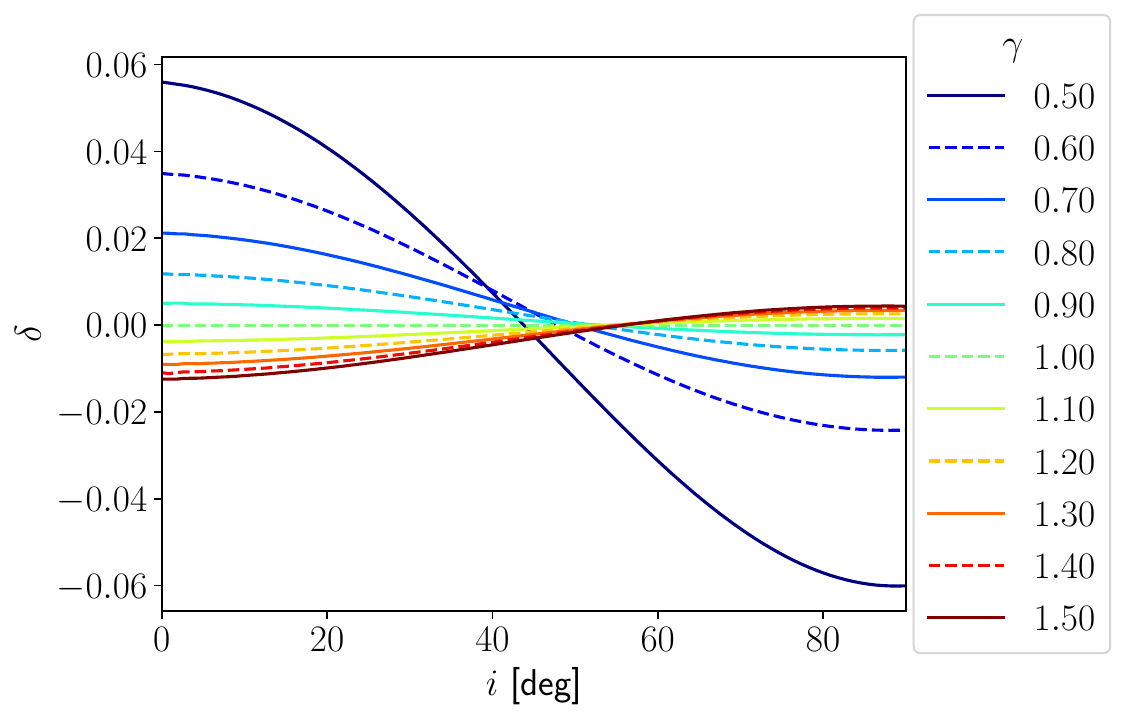}
    \caption{Same as Fig.~\ref{kerr_sh} for the $\gamma$-metric ($\gamma=0.50$--$1.50$).}
    \label{gamma_sh}
\end{figure*}

First, let us examine how the shadow appears from different viewing angles. Due to the axial symmetry of the spacetime, we can study the effects of the black hole spin and deformation parameters on the shadow. The characteristic areal radius of the shadow curve, $r_{\mathrm{sh}}$, and the Schwarzschild Deviation Parameter, $\delta$, are shown in Figs.~\ref{kerr_sh}--\ref{gamma_sh}.

The areal radius $r_{\mathrm{sh}}$ is defined as
\begin{equation}
r_{\mathrm{sh}} \equiv \sqrt{\frac{A}{\pi}},
\end{equation}
where $A$ is the total area enclosed by the shadow curve.

The parameter $\delta$ quantifies the fractional deviation of the shadow size compared to the Schwarzschild case, and is defined as
\begin{equation}
\delta \equiv \frac{r_{\mathrm{sh}}}{r_{\mathrm{sch}}} - 1,
\end{equation}

\noindent where $r_{\mathrm{sch}}=\sqrt{27}\,M$ is the shadow radius of a Schwarzschild 
black hole for an observer at infinity. 
In numerical ray-tracing we approximate this limit by placing the observer’s 
screen at a large but finite distance $d = 10^5 M$, which reproduces the 
observer-at-infinity result to high accuracy.

Figure~\ref{kerr_sh} shows the behavior of $r_{\mathrm{sh}}$ and $\delta$ for the Kerr spacetime. As the spin parameter $a_*$ increases, the shadow areal radius $r_{\mathrm{sh}}$ gradually decreases. The maximum deviation from the Schwarzschild case occurs when the observer’s inclination angle is $0^\circ$ (the observer is along the black hole spin axis) and the black hole spin is near its maximum. Even in this extreme case, the deviation remains below approximately $7.5\%$~\cite{Takahashi_2004, EHT_SgrA_VI}. Similarly, Fig.~\ref{gamma_sh} illustrates how variations in the parameter $\gamma$ of the $\gamma$--metric modify both the shadow radius and the Schwarzschild deviation. In this case, the deformation leads to deviations in $\delta$ ranging approximately from $-0.07$ to $0.06$.

Observational data from the Keck and VLTI instruments provide important constraints on the shadow parameters of Sgr~A*. 
The Event Horizon Telescope (EHT) Collaboration~\cite{EHT_SgrA_I} reported a ring image of Sgr~A* with a diameter of $51.8 \pm 2.3\ \mu\mathrm{as}$ at a wavelength of 1.3 mm in the $1\sigma$ level. Assuming a distance of $D = 8.15 \pm 0.15$~kpc~\cite{EHT_SgrA_IV} and the observed object as a Schwarzschild black hole, this angular diameter corresponds to an estimated black hole mass of $M = 4.0^{+1.1}_{-0.6} \times 10^6 M_\odot$.

Distance and mass were independently estimated by the Keck~\cite{Keck_2019} and VLTI~\cite{VLTI_2021,VLTI_2022} observations as
\[
D = 7935 \pm 50 \pm 32\ \mathrm{pc}, \;\;  M = (3.951 \pm 0.047) \times 10^6 M_\odot \quad (\text{Keck}),
\]
\[ D = 8277 \pm 9 \pm 33\ \mathrm{pc}, \;\; M = (4.297 \pm 0.013) \times 10^6 M_\odot \quad (\text{VLTI}). \].

The Schwarzschild deviation parameter $\delta$ was constrained from the EHT observations by comparison with the VLTI and Keck measurements as
\[
\delta = -0.04^{+0.09}_{-0.10} \quad (\text{Keck}), \qquad \delta = -0.08 \pm 0.09 \quad (\text{VLTI}).
\]

The corresponding bounds on the shadow radius $r_{\mathrm{sh}}$ are approximately:
\begin{align*}
4.47 \; M &\lesssim r_{\mathrm{sh}} \lesssim 5.46 \; M \quad (\text{Keck}),\\
4.31 \; M &\lesssim r_{\mathrm{sh}} \lesssim 5.25 \; M \quad (\text{VLTI}).
\end{align*}

\begin{figure*}[ht!]
    \centering
    \includegraphics[width=0.49\linewidth]{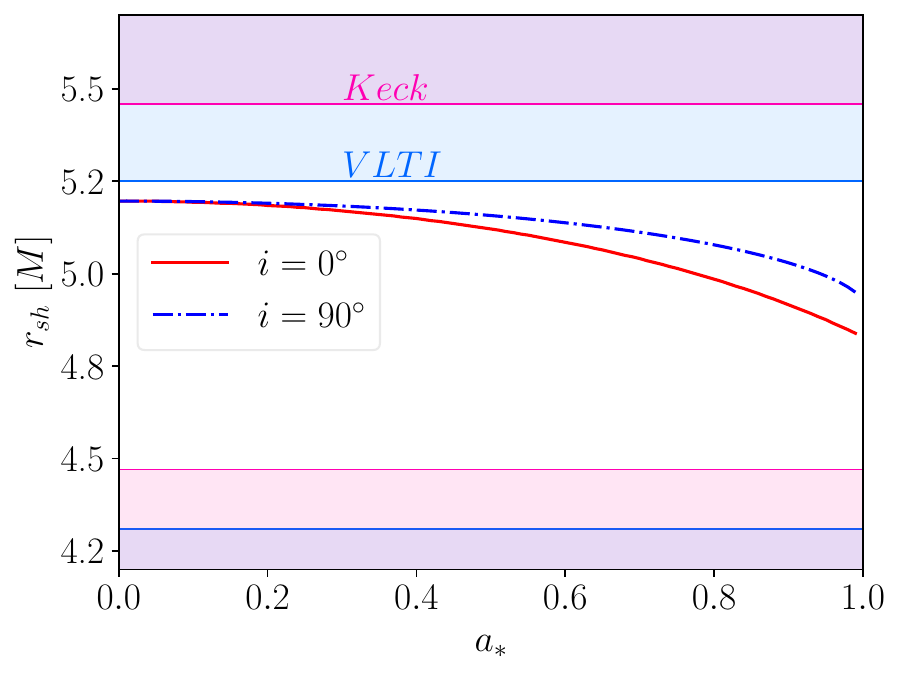}
    \includegraphics[width=0.49\linewidth]{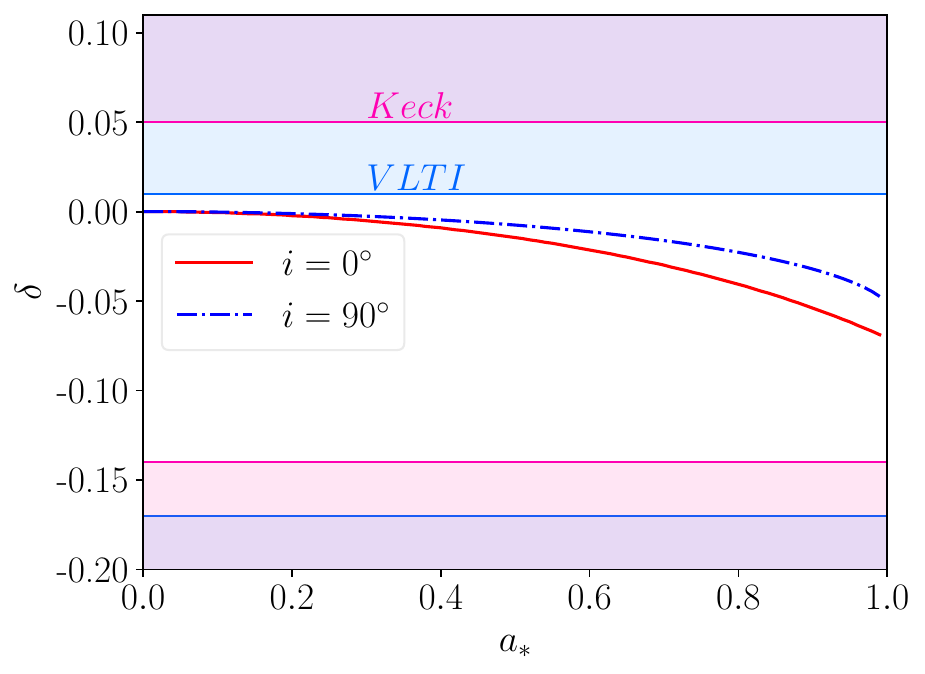}
    \caption{Sgr~A* Kerr shadow constraints: $r_{\rm sh}$ (left) and Schwarzschild deviation parameter $\delta$ (right) vs. spin $a_*$ for $i=0^\circ$ (red solid) and $90^\circ$ (blue dash--dotted). Shaded pink and light-blue bands indicate VLTI and Keck limits.}
    \label{kerr_const}
\end{figure*}

\begin{figure*}[ht!]
    \centering
    \includegraphics[width=0.49\linewidth]{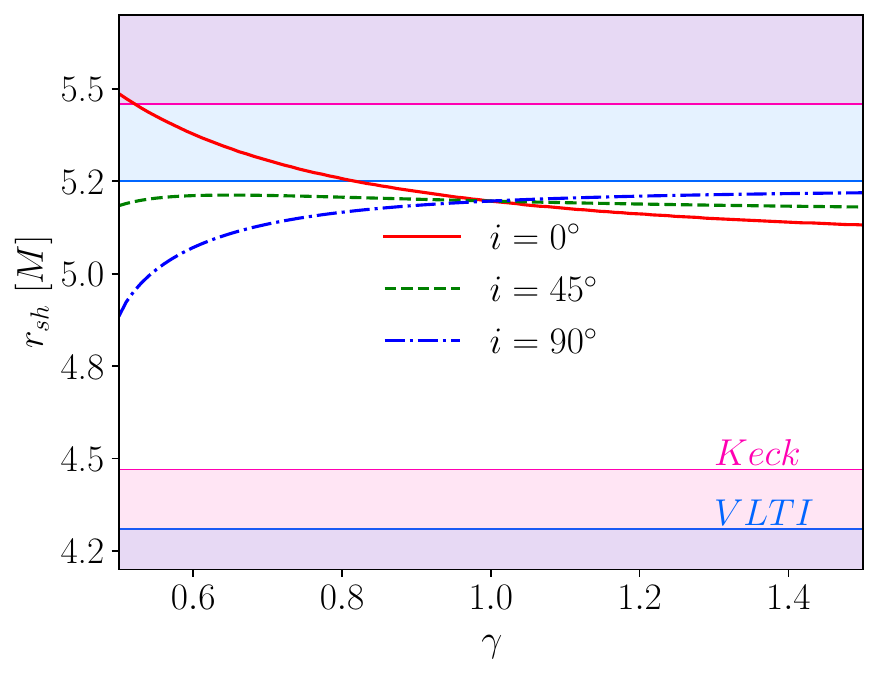}
    \includegraphics[width=0.49\linewidth]{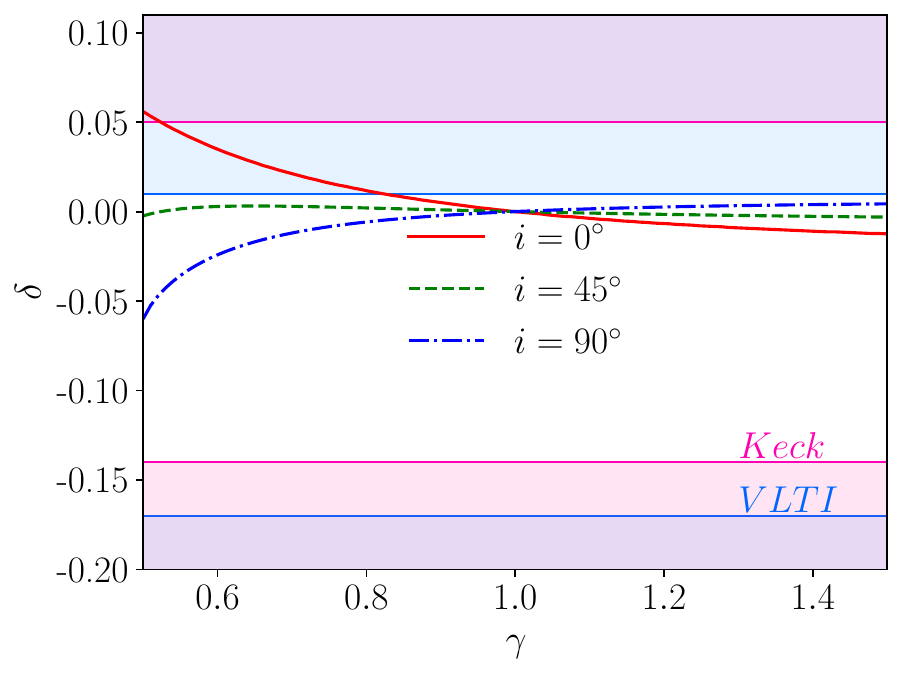}
    \caption{Same as Fig.~\ref{kerr_const} for the $\gamma$ metric at $i=0^\circ,\,45^\circ,\,90^\circ$.}
    \label{gamma_const}
\end{figure*}

\begin{figure*}[ht!]
    \centering
    \includegraphics[width=0.49\linewidth]{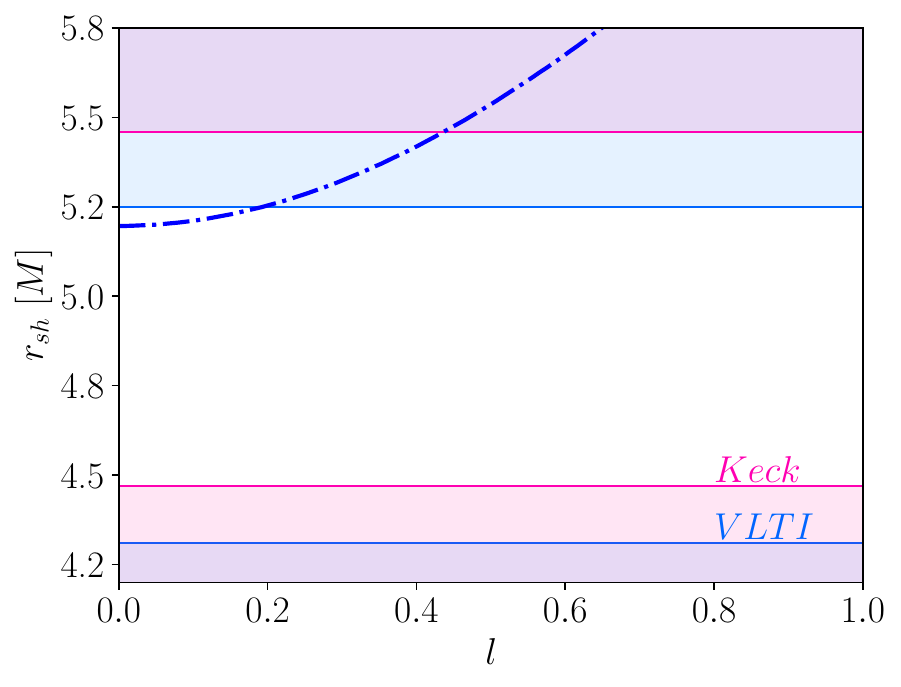}
    \includegraphics[width=0.49\linewidth]{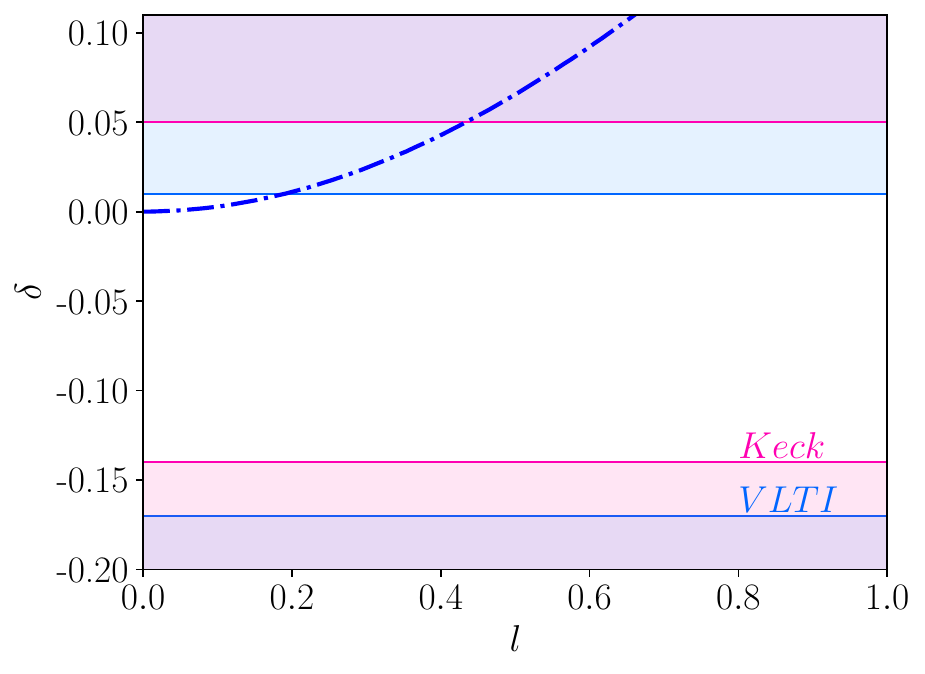}
    \caption{Same as Fig.~\ref{kerr_const} for the Taub-NUT metric: $r_{\rm sh}$ and $\delta$ vs. NUT parameter $l$.}
    \label{nut_const}
\end{figure*}

\begin{figure*}[ht!]
    \centering
    \includegraphics[width=0.49\linewidth]{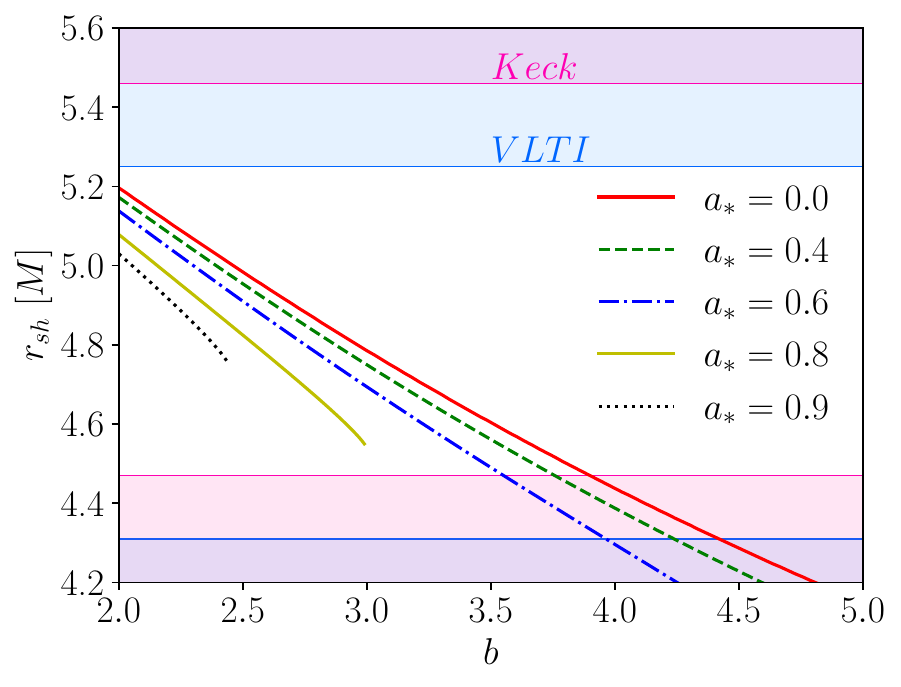}
    \includegraphics[width=0.49\linewidth]{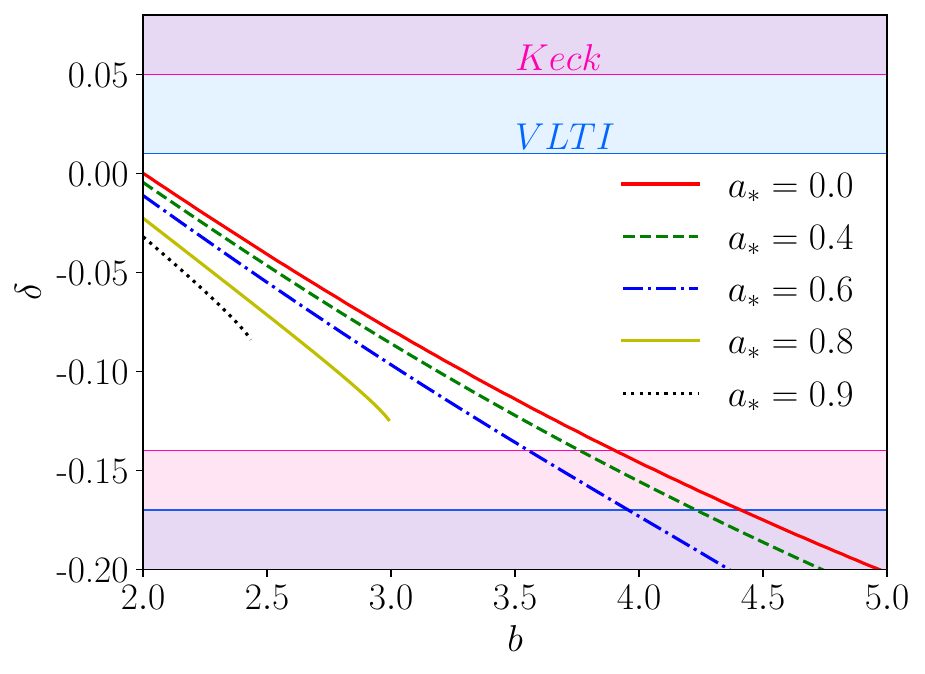}
    \caption{Same as Fig.~\ref{kerr_const} for the Kaluza-Klein metric: $r_{\rm sh}$ and $\delta$ vs. parameter $b$ for $a_*=0.0$--$0.9$.}
    \label{kal_const}
\end{figure*}

These observational constraints are shown as shaded bands in Figs.~\ref{kerr_const}--\ref{kal_const}, comparing Kerr, $\gamma$, Taub-NUT, and Kaluza-Klein metrics. For the Kerr case (Fig.~\ref{kerr_const}), no constraint on the black hole spin is obtained, as all computed shadows lie within the observationally allowed regions for both Keck and VLTI data. In the $\gamma$ metric (Fig.~\ref{gamma_const}), a similar situation occurs for most inclination angles. However, for an observer close to the symmetry axis viewing angles ($i = 0^\circ$), the parameter $\gamma$ is constrained: $\gamma > 0.52$ from Keck data and $\gamma > 0.82$ from VLTI data.

For the Taub-NUT metric (Fig.~\ref{nut_const}), observational data place an upper limit on the NUT parameter $l$: it must satisfy $0 \leq l < 0.44$ according to Keck and $0 \leq l < 0.2$ according to VLTI. 

In the Kaluza-Klein metric (Fig.~\ref{kal_const}), considering the case where the magnetic and electric charges are equal ($b = c$), the allowed range for the parameter $b$ is $2 \leq b < 3.9$ for Keck and $2 \leq b < 4.42$ for VLTI observations.

\section{Shadow Characterization: Coordinate-Independent Formalism}
\label{sec:char}

An accurate characterization of black hole shadows is crucial for constraining the properties of black holes. In this section, we explore two mathematical approaches for describing black hole shadows: the Legendre expansion and the Fourier expansion. Both expansions allow for the representation of the shadow in a coordinate-independent manner. However, we will demonstrate that the Fourier expansion offers superior flexibility, especially when rotational effects are explored. The Fourier expansion is ideal for signal processing, error analysis, especially in periodic systems, etc.

\begin{figure}
    \centering
    \includegraphics[width=0.6\linewidth]{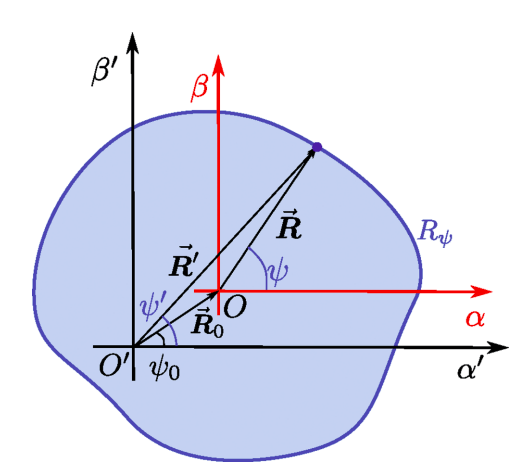}
    \caption{Diagram of the black hole shadow rendered as the polar curve~$R_\psi$ in the $(\alpha,\beta)$ plane with origin~$O$ at the shadow’s center, which appears shifted by the vector~$\mathbf{R}_0$ in the observer’s frame $(\alpha',\beta')$. Adopted from ~\cite{Abdujabbarov_15}.}

    \label{fig:schematic}
\end{figure}


The black hole shadow is typically represented as a polar curve $R(\psi)$, which can be defined in an arbitrary coordinate system. A schematic example of such a representation is shown in Fig.~\ref{fig:schematic}, where the shadow appears as a closed curve centered at $O$, while the observer’s coordinate origin $O'$ may be offset. To describe the shadow in a coordinate-independent way, it is useful to compute the effective center of the shadow, denoted by $R_0$, using the following integrals~\cite{Abdujabbarov_15}:

\begin{equation}
\begin{aligned}
R_{0}:=&\left(\int_{0}^{2 \pi} R^{\prime} \mathrm{d} \psi^{\prime}\right)^{-1}\left[\left(\int_{0}^{2 \pi} R^{\prime 2} \cos \psi^{\prime} \mathrm{d} \psi^{\prime}\right)^{2}\right.\\
&\left.+\left(\int_{0}^{2 \pi} R^{\prime 2} \sin \psi^{\prime} \mathrm{d} \psi^{\prime}\right)^{2}\right]^{1 / 2}\ .
\end{aligned} \label{eq:center_mass}
\end{equation}

Equation~(\ref{eq:center_mass}) defines the effective center $R_0$ of the shadow in a 
coordinate-independent way. The motivation for this construction is that 
the geometric center of the observed contour does not necessarily coincide 
with the observer’s coordinate origin $O'$. Simply expanding $R'(\psi')$ 
around $O'$ could therefore introduce spurious asymmetries. The integral 
expression for $R_0$ plays the role of a centroid, analogous to the 
center of mass of a closed curve, and provides a robust way to anchor the 
shadow’s effective center before carrying out further analysis.

The angular displacement of the center of the shadow, $\psi_0$, is then given by~\cite{Abdujabbarov_15}:

\begin{equation}
\psi_0 := \tan^{-1} \left( \frac{\int_{0}^{2\pi} R^{\prime 2} \sin \psi' \, d\psi'}{\int_{0}^{2\pi} R^{\prime 2} \cos \psi' \, d\psi'} \right)\ .
\end{equation}

A new coordinate system can be defined, where the center is moved to $R_0$ and the shadow is transformed into the coordinates $(R, \psi)$ as follows~\cite{Abdujabbarov_15}:

\begin{equation}
\begin{aligned}
R:=& {\left[\left(R^{\prime} \cos \psi^{\prime}-R_{0} \cos \psi_{0}\right)^{2}\right.} \\
&\left.+\left(R^{\prime} \sin \psi^{\prime}-R_{0} \sin \psi_{0}\right)^{2}\right]^{1 / 2}, \\
\psi:=& \tan ^{-1} \frac{R^{\prime} \sin \psi^{\prime}-R_{0} \sin \psi_{0}}{R^{\prime} \cos \psi^{\prime}-R_{0} \cos \psi_{0}  }.
\end{aligned} 
\end{equation}

The Legendre expansion is a method that is used to represent the shadow of a black hole as a series of Legendre polynomials. It is particularly useful when the shadow is symmetric in nature and is observed from a fixed inclination. The general form of the Legendre expansion is given by~\cite{Abdujabbarov_15}:

\begin{equation}
R(\psi) = \sum_{\ell=0}^{\infty} c_\ell P_\ell(\cos \psi)\ , 
\end{equation}

where \( R(\psi) \) is the radial distance of the shadow at an angle \( \psi \), \( P_\ell(\cos \psi) \) are the Legendre polynomials of order \( \ell \), and the coefficients \( c_\ell \) capture the contribution from each polynomial order.

Then the expansion coefficients $c_\ell$'s have been calculated as~\cite{Abdujabbarov_15}
\begin{equation*}
c_{\ell} =\frac{2\ell+1}{2}
\int_{\lambda_1}^{\lambda_2} R(\lambda)
P_{\ell}(\cos\psi) \sin\psi  \left(\frac{d\psi}{d\lambda}\right)
d\lambda\ , \label{ancells}
\end{equation*}
for the values of $\ell =0,\ ..., \ 5$.

Another effective approach to represent the shadow is through a Fourier series expansion of the shadow radius \( R(\psi) \). The general form of the Fourier expansion is as follows:

\begin{equation}
R(\psi) = a_0 + \sum_{\ell=1}^{\infty} \left[ a_\ell \cos(\ell\psi) + b_\ell \sin(\ell\psi) \right]\ .
\end{equation}

The Fourier coefficients \(a_\ell\) and \(b_\ell\) are calculated by integrating the shadow function over the angular variable \( \psi \):

\begin{equation}
a_0 = \frac{1}{2\pi} \int_0^{2\pi} R(\psi) \, d\psi\ ,
\end{equation}
\begin{equation}
a_\ell = \frac{1}{\pi} \int_0^{2\pi} R(\psi) \cos(\ell\psi) \, d\psi\ ,
\end{equation}
\begin{equation}
b_\ell = \frac{1}{\pi} \int_0^{2\pi} R(\psi) \sin(\ell\psi) \, d\psi\ .
\end{equation}

Additionally, the modulus (or magnitude) of the Fourier coefficients is defined as:

\begin{equation}
C_0 = a_0, \quad C_\ell = \sqrt{a_\ell^2 + b_\ell^2} \quad \text{for} \quad \ell \ge 1\ .
\end{equation}

The use of the modulus (or magnitude) of the Fourier coefficients \( C_\ell \) offers a key advantage in characterizing black hole shadows: these quantities are invariant under coordinate rotation. This property is particularly beneficial when comparing shadows across different spacetimes, as it ensures that the results are independent of the orientation of the shadow in the observer's image plane.

To illustrate this invariance, consider a rotation of the shadow curve \( R(\psi) \) by an angle \( \psi_0 \), such that the rotated curve becomes \( R_{\psi_0}(\psi) = R(\psi + \psi_0) \). The new Fourier coefficients transform as follows:
\begin{align*}
a_\ell' &= \cos(\ell \psi_0)\, a_\ell + \sin(\ell \psi_0)\, b_\ell, \\
b_\ell' &= -\sin(\ell \psi_0)\, a_\ell + \cos(\ell \psi_0)\, b_\ell,
\end{align*}
leading to the invariance of their Euclidean norm:
\[
C_{\ell}' = \sqrt{a_\ell'^2 + b_\ell'^2} = \sqrt{a_\ell^2 + b_\ell^2} = C_\ell.
\]

This demonstrates that \( C_\ell \) remains unchanged under rotations, and thus serves as a robust, rotation-invariant descriptor of the shadow curve. Such invariance makes the set \( \{C_\ell\} \) an ideal tool for shadow comparison, model fitting, and parameter inference in a coordinate-independent manner.

\begin{figure*}[ht!]
    \centering
    \includegraphics[width=0.49\linewidth]{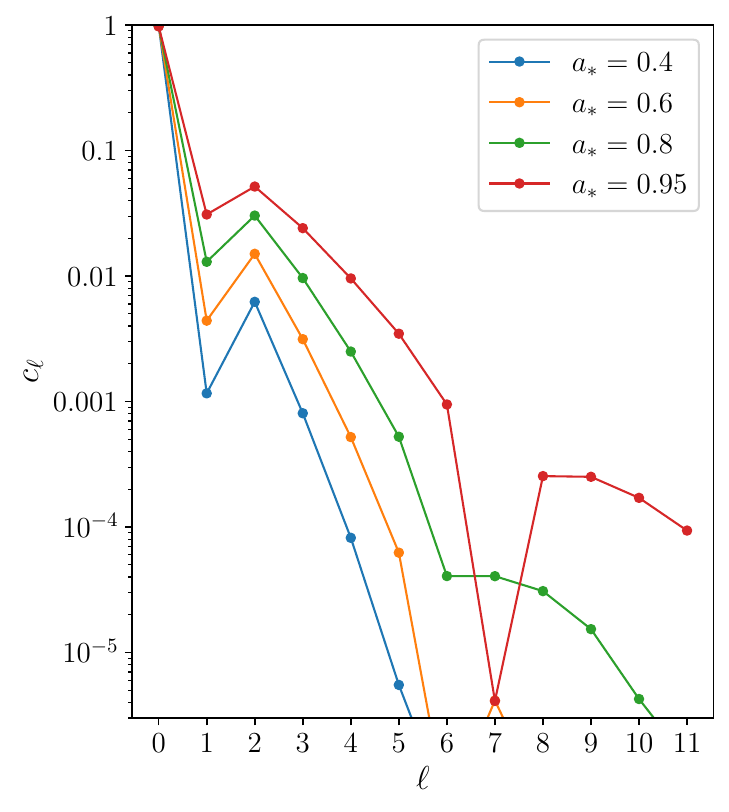}
     \includegraphics[width=0.49\linewidth]{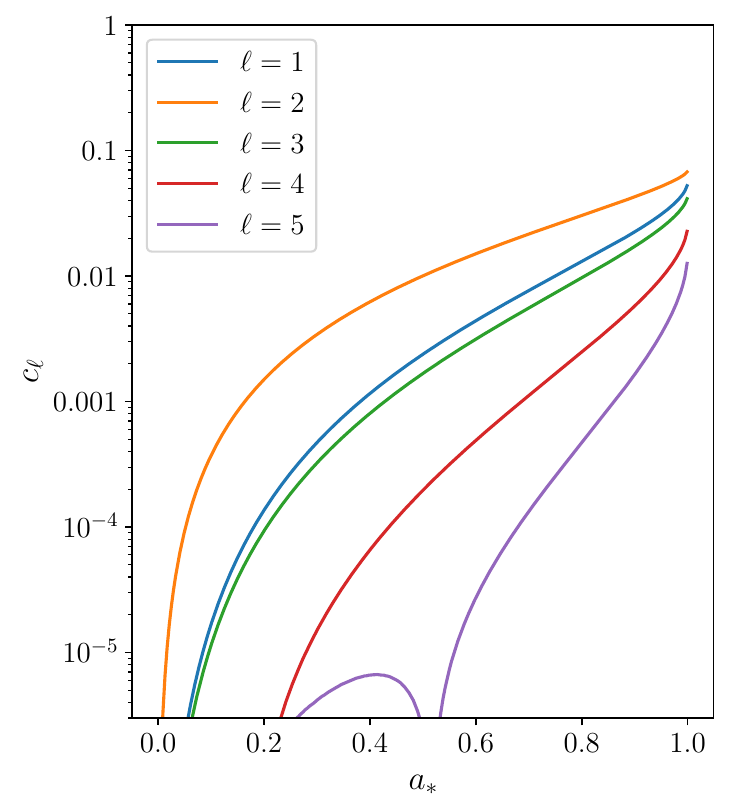}

    \includegraphics[width=0.49\linewidth]{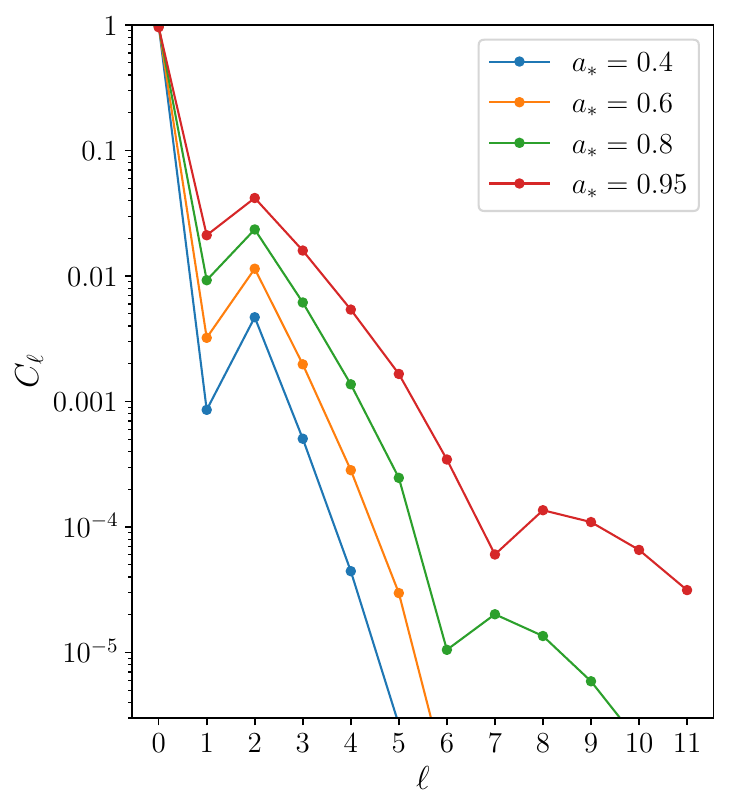}
     \includegraphics[width=0.49\linewidth]{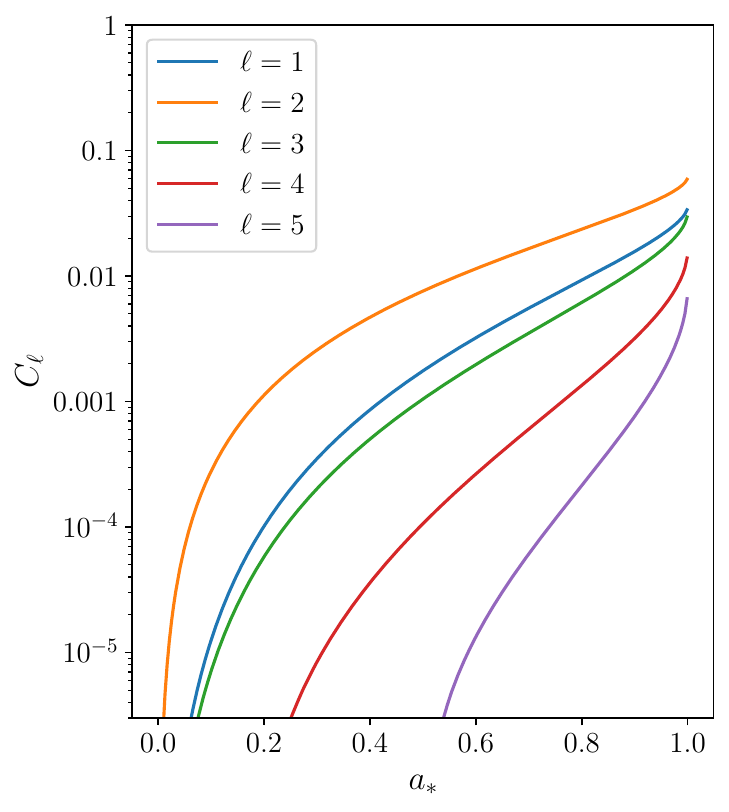}
     
    \caption{Behavior of the shadow expansion magnitudes $c_\ell$ ($C_\ell$) illustrating their rapid decay. Top left: Legendre expansion magnitudes versus mode order $\ell$ for spins $a_*=0.4,0.6,0.8,0.95$. Top right: Legendre expansion magnitudes versus $a_*$ for $\ell=1–5$. Bottom panels: the corresponding Fourier expansion amplitudes plotted in the same two formats.}

    \label{coord_ind}
\end{figure*}

To evaluate the utility of the expansion methods, we generate approximately \( N = 10^5 \) Kerr black hole shadow images by varying the spin parameter \( a_* \in (0, 1) \) and inclination angle \( i \in (0, \pi/2) \). For each image, we compute both Legendre and Fourier coefficients. Figure~\ref{coord_ind} shows the behavior of the expansion magnitudes \( C_\ell \), demonstrating their rapid decay. The top panels display Legendre coefficients as a function of mode order \( \ell \) (left) and spin \( a_* \) (right), while the bottom panels show the corresponding Fourier coefficients. Both expansions converge quickly, with higher-order terms contributing less for lower spin values. This confirms that only a few modes are necessary to reconstruct the shadow curve with high accuracy.

\begin{figure*}[ht!]
    \centering
    \includegraphics[width=0.49\linewidth]{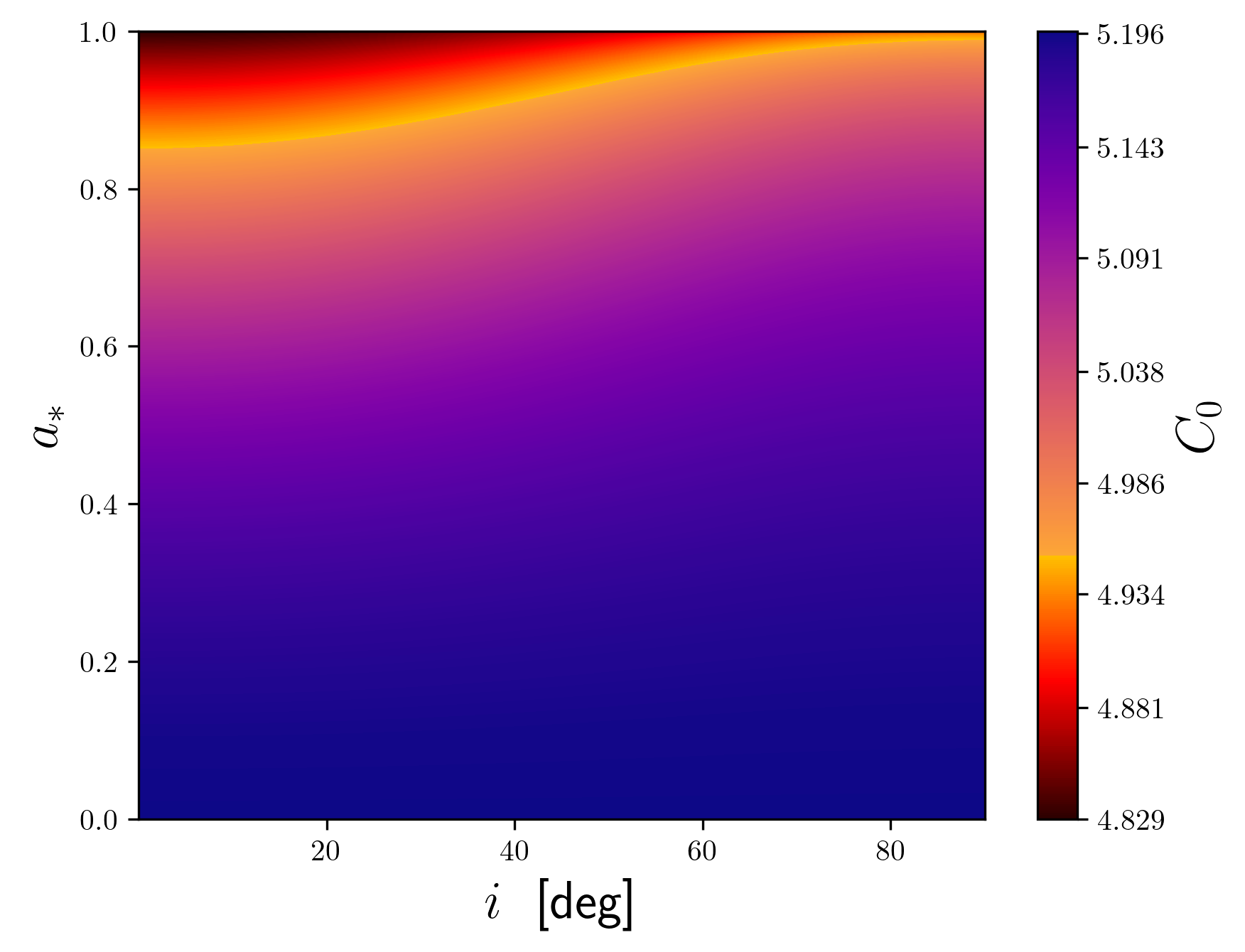}
     \includegraphics[width=0.49\linewidth]{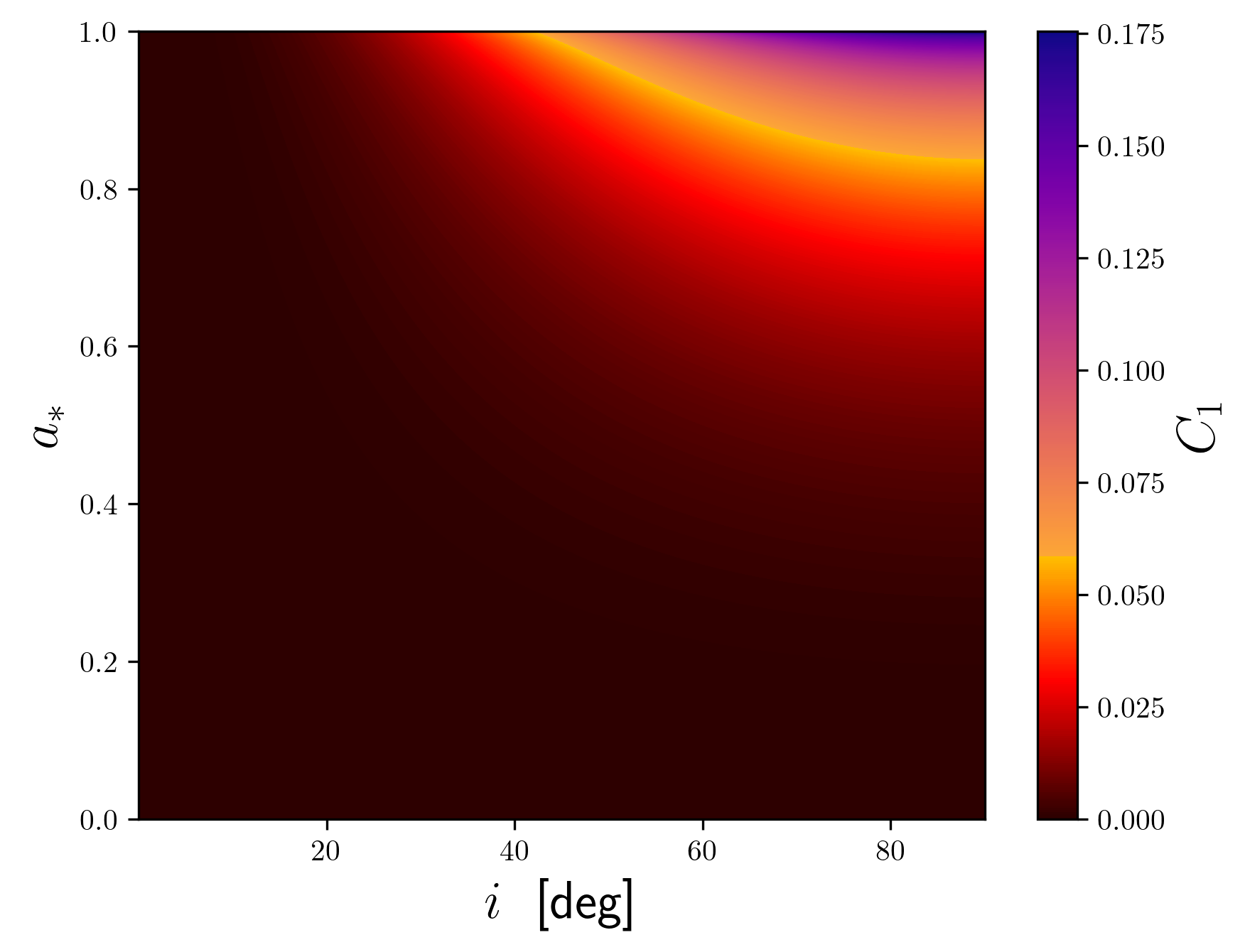}

    \includegraphics[width=0.49\linewidth]{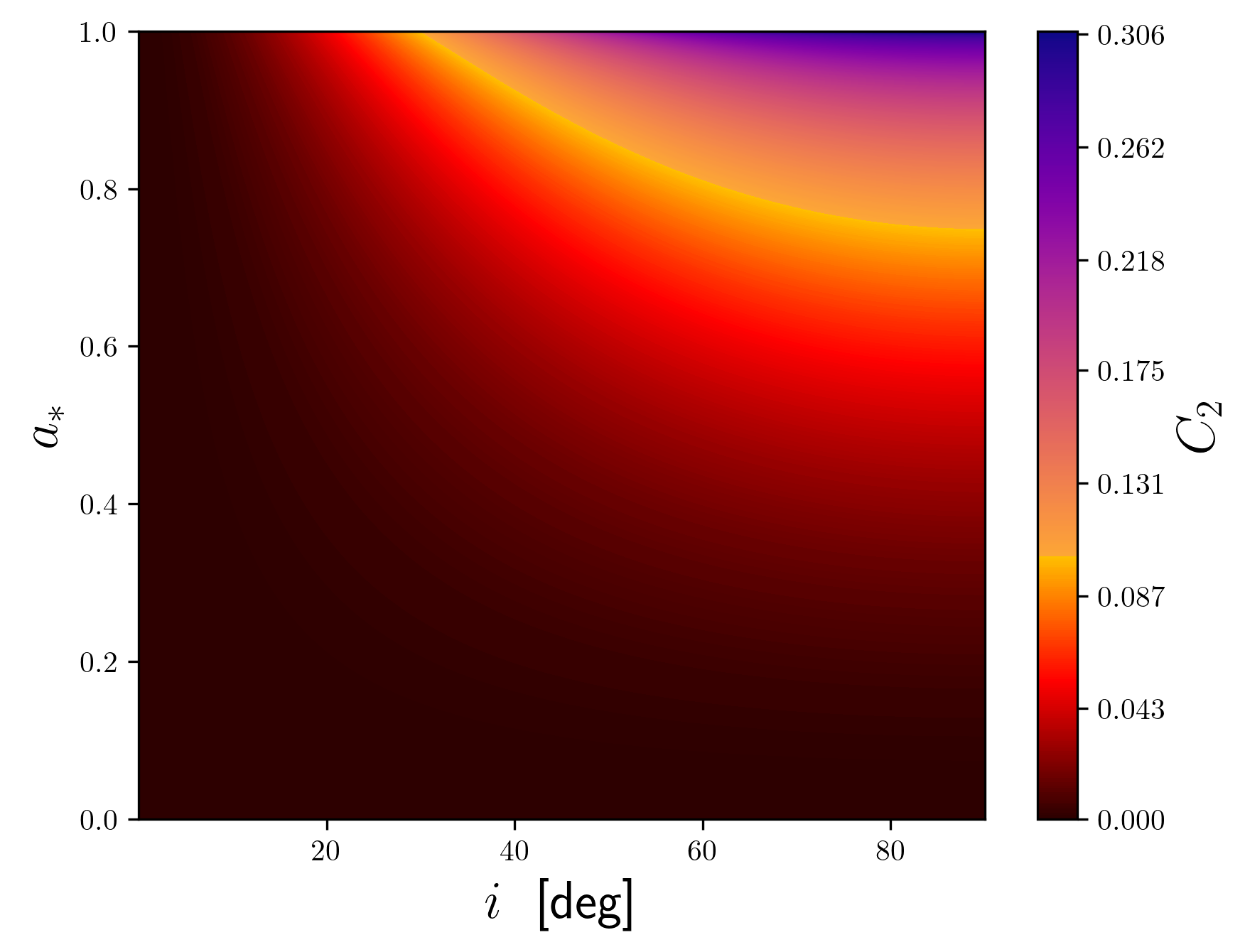}
     \includegraphics[width=0.49\linewidth]{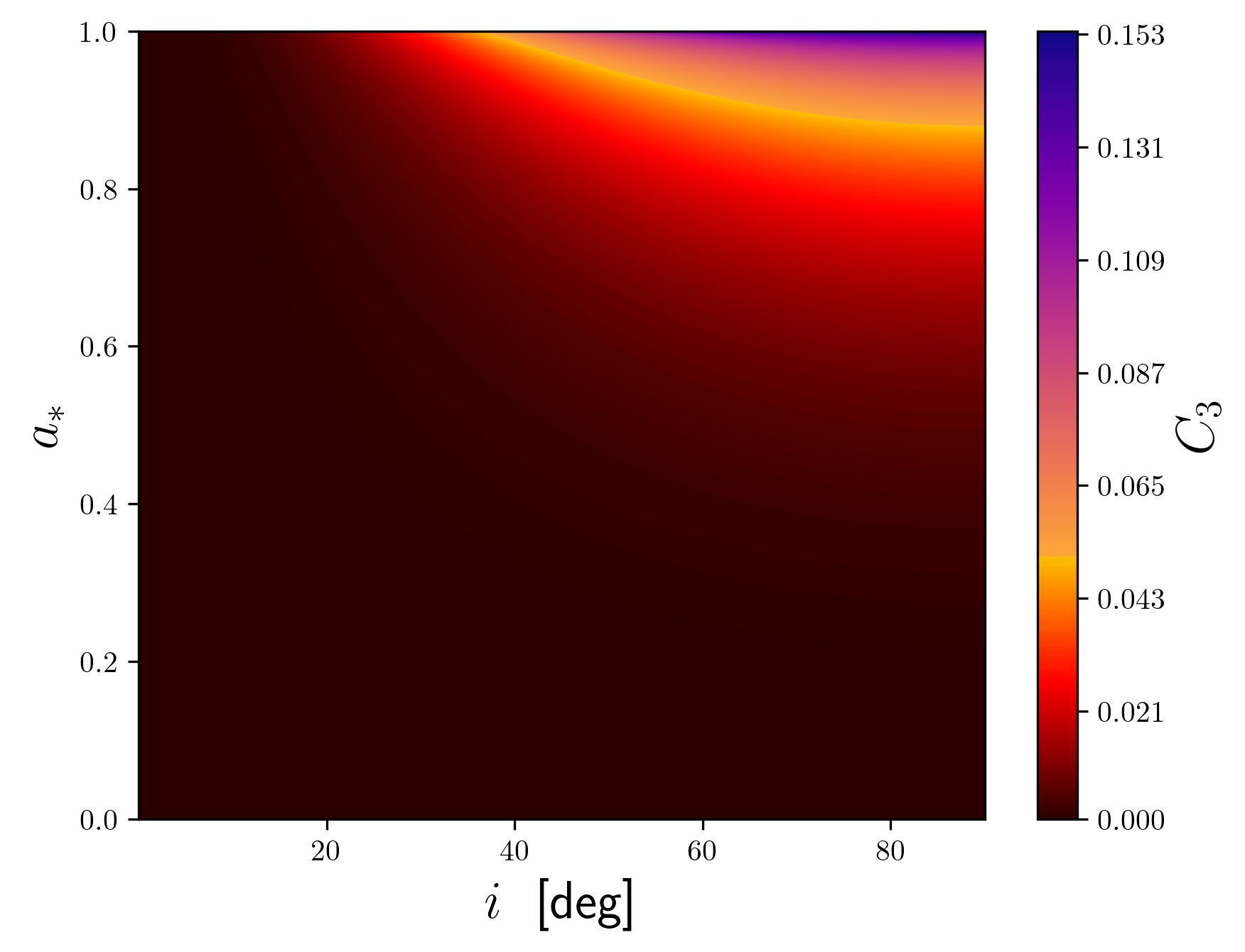}
    \caption{Heatmaps of the first four Fourier coefficients $C_0$–$C_3$ for a Kerr black hole shadow, shown as functions of inclination $i$ and spin $a_*$ in the panels (top left to bottom right).}

    \label{coord_ind_map}
\end{figure*}

In addition, Figure~\ref{coord_ind_map} presents heatmaps of the first four Fourier coefficients \( C_0 \)–\( C_3 \), plotted over the parameter space of spin \( a_* \) and inclination angle \( i \). The coefficient \( C_0 \) reflects the average size of the shadow and varies monotonically, while the higher-order coefficients encode features related to asymmetry and distortion. These patterns reveal how Fourier coefficients contain rich, structured information that can be used for inverse parameter estimation from observational shadow data.

\section{Neural Network Framework}
\label{sec:nn}

To establish a bidirectional mapping between black hole parameters and their corresponding shadow morphology, we develop a two-stage machine learning framework based on neural networks. The first neural network predicts the shadow curve from physical parameters, while the second one inversely recovers the parameters from observed shadow features using Fourier coefficients. The coordinate independent formalism, introduced in Section~\ref{sec:char}, is specifically employed in the second network to ensure robust and rotation-invariant parameter estimation.

We construct the neural networks using the rotating $\gamma$ spacetime (see Appendix \ref{metric-delta}) as a representative example. The physical parameters relevant to shadow formation include spin parameter \( a_* \), mass \( M \) of the black hole, and inclination angle \( i \), which corresponds to the observer’s viewing direction relative to the spin axis of the black hole. In addition, we consider a deformation parameter \( \gamma \), which characterizes deviations from Kerr geometry while remaining within the framework of general relativity. 
Since the overall size of the black hole shadow scales with mass, the shadow radius is normalized by \( M \).

Shadow profiles are generated numerically using a ray tracing algorithm, as described in Section~\ref{sec:shadows}, since analytical expressions for the shadow boundary are generally unavailable in deformed Kerr spacetime. To train neural networks, we produce more than one million synthetic shadow images by uniformly sampling the parameter space: spin \( a_* \in [0, 1] \), inclination angle \( i \in [0^\circ, 90^\circ] \), and deformation parameter \( \gamma \in [0.5, 1.5] \).

\subsection{Forward Network: Shadow Generation from Parameters}

The first model is designed as a regression neural network that predicts the black hole shadow shape from its physical parameters. Specifically, it maps the spin \( a_* \), the inclination angle \( i \), and the deformation parameter \( \gamma \) to the corresponding polar shadow function \( R(\psi) \). To make the problem tractable for training, the function \( R(\psi) \) is discretized into \( N_\psi = 360 \) uniformly spaced angular bins over \( \psi \in [0, 2\pi] \), representing the radial distances along the shadow boundary.

The network architecture consists of:
\begin{itemize}
    \item \textbf{Input layer}: parameters \( a_* \), \( i \), and \( \gamma \),
    \item \textbf{Hidden layers}: three fully connected layers with 256, 512, and 256 neurons, each using ReLU activation,
    \item \textbf{Output layer}: 360 values corresponding to the discretized shadow profile \( R(\psi) \).
\end{itemize}

The network is trained using the mean squared error (MSE) as the loss function. The final training converges to an average MSE of approximately \( 6.7 \times 10^{-6} \), indicating a high degree of accuracy in reconstructing the shadow contour.

\begin{figure}[ht]
    \centering
    \includegraphics[width=0.45\textwidth]{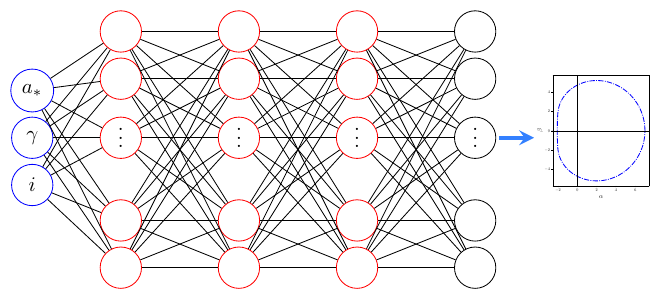}
    \caption{Schematic diagram of the forward neural network architecture used to predict the shadow profile \( R(\psi) \) from input parameters \( (a_*, i, \gamma) \).}
    \label{f-NN_diag1}
\end{figure}

\begin{figure}[ht]
    \centering
    \includegraphics[width=0.45\textwidth]{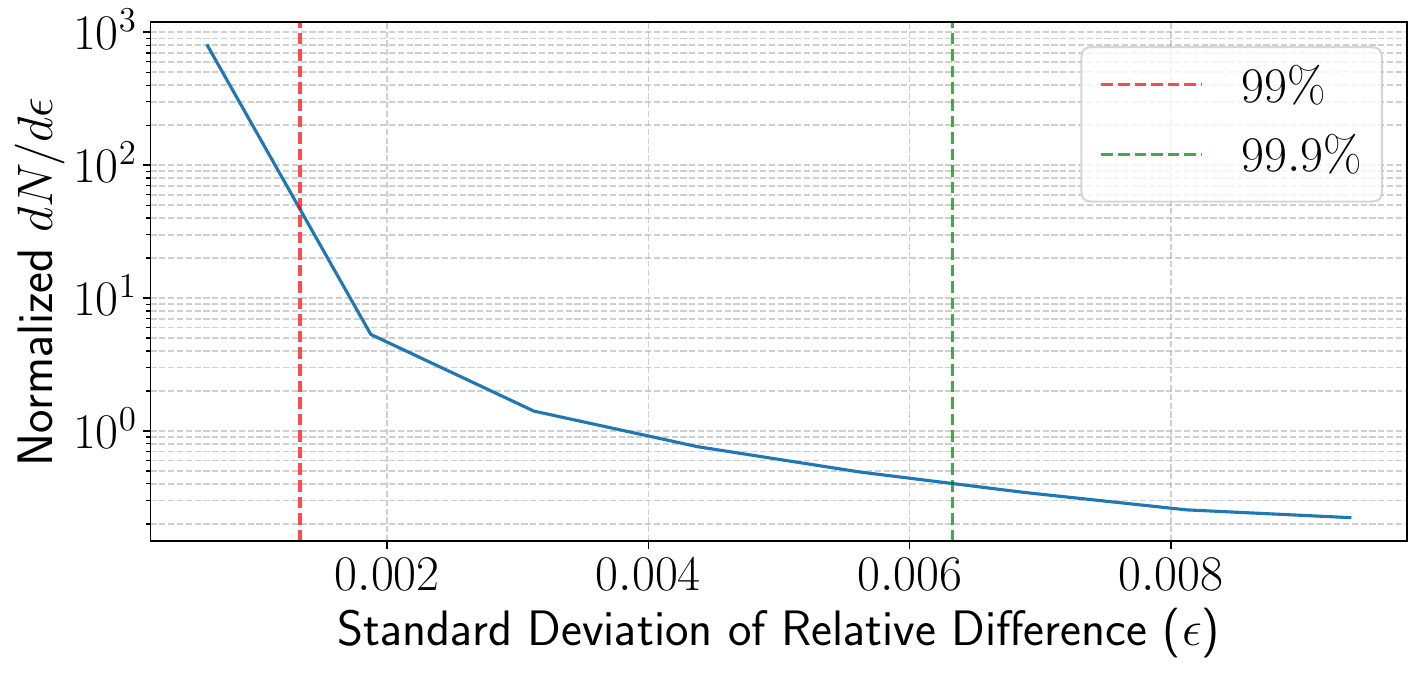}
    \caption{Normalized distribution \( dN/d\epsilon \) as a function of the standard deviation \( \epsilon \) of the relative error between predicted and true shadow profiles.}
    \label{f-epsilon_distribution}
\end{figure}

Figure~\ref{f-NN_diag1} provides a schematic representation of the network architecture used in the forward model. In addition, Figure~\ref{f-epsilon_distribution} shows the distribution \( dN/d\epsilon \), where \( \epsilon \) represents the standard deviation of the relative error between the predicted and ground truth shadows. This plot confirms that the prediction errors remain low and well controlled throughout the dataset.

Finally, although the current model is trained specifically using the rotating $\gamma$ spacetime, the framework is general and can be extended to other black hole metrics by regenerating the training data accordingly.

It is important to note that we did not attempt to predict the Fourier coefficients  $C_\ell$ directly from the physical parameters. Instead, the forward network is constructed to generate the full shadow profile $R(\psi)$, which mimics the type of data provided by observations. From these synthetic contours, the Fourier coefficients can then be extracted and used as inputs to the inverse network for parameter estimation. This two-step design ensures consistency with the observational pipeline, where images are first reconstructed and only afterward decomposed into invariant features.

\subsection{Inverse Network: Parameter Estimation from Shadow Features}

The second neural network is trained to estimate black hole parameters from the shadow shape. Instead of using the full function \( R(\psi) \), we use a small set of normalized Fourier magnitudes:
\[
\left\{ C_0, \frac{C_1}{C_0}, \frac{C_2}{C_0}, \dots, \frac{C_n}{C_0} \right\}.
\]
These values are invariant under rotation and help the network stay accurate even when the shadow is rotated.

The model learns to predict the spin \( a_* \), the inclination angle \( i \), and the deformation parameter \( \gamma \). 
The network architecture consists of:
\begin{itemize}
    \item \textbf{Input layer}: 5 normalized Fourier coefficients (for \( n = 4 \)),
    \item \textbf{Hidden layers}: four fully connected layers with 128, 256, 128, and 64 neurons, each using ReLU activation,
    \item \textbf{Output layer}: three values for \( a_*, i, \gamma \), scaled between 0 and 1 using a Sigmoid activation.
\end{itemize}

The inverse network is trained on the same dataset used in the forward model, which was generated by ray tracing in the rotating $\gamma$ spacetime over uniformly sampled parameter ranges.

The model is trained using mean squared error (MSE) loss. The final results for different Fourier orders are:
\begin{itemize}
    \item For \( n = 2 \): MSE \( \approx 3.89 \times 10^{-3} \),
    \item For \( n = 3 \): MSE \( \approx 9.48 \times 10^{-4} \),
    \item For \( n = 4 \): MSE \( \approx 4.33 \times 10^{-4} \).
\end{itemize}

\begin{figure}[ht]
    \centering
    \includegraphics[width=0.45\textwidth]{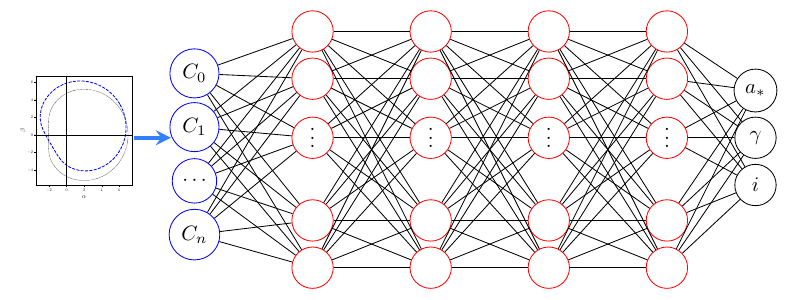}
    \caption{Architecture of the inverse neural network. The model takes five normalized Fourier coefficients as input and predicts the black hole parameters: spin \( a_* \), inclination \( i \), and deformation \( \gamma \). }
    \label{f-NN_diag2}
\end{figure}

\begin{figure*}[ht!]
    \centering
    \includegraphics[width=0.9\linewidth]{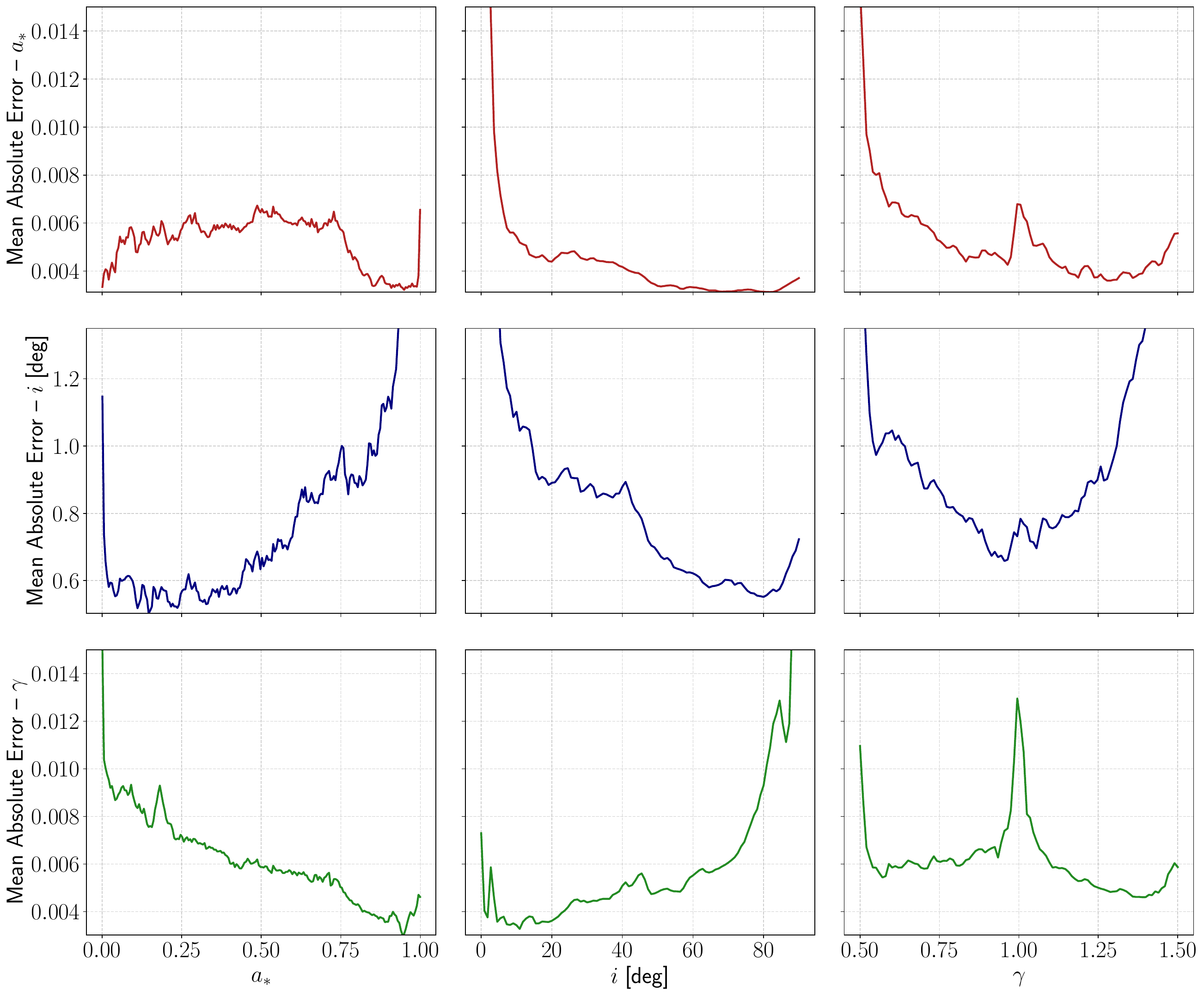}
    \caption{Mean absolute relative error analysis for the inverse neural network. Each subplot shows how the prediction error varies for spin \( a_* \), inclination angle \( i \), and deformation parameter \( \gamma \), while averaging over the other two parameters. Inclination errors are shown in degrees. The 3×3 grid highlights regions of higher or lower accuracy across the input space.}
    \label{f-NN_error2}
\end{figure*}

Figure~\ref{f-NN_diag2} shows the network architecture. In addition, 
Figure~\ref{f-NN_error2} shows the mean absolute relative error for each predicted parameter across the input space. Each subplot varies one parameter while averaging over the others, allowing a clear view of how the accuracy of the network depends on spin, inclination, and deformation. Inclination errors are shown in degrees. This visualization highlights that the model performs consistently well across most of the domain, with slightly larger errors near edge cases. Most of the prediction errors occur in regions of the parameter space where the shadow shape changes very little with respect to variations in the input. For example, when the spin is low and the deformation parameter \( \gamma = 1 \) (corresponding to the Kerr case), changing the inclination angle has only a minor effect on the shadow curve. As a result, the network struggles to distinguish between different inclinations and may also confuse similar configurations for spin and \( \gamma \). These degeneracies make accurate predictions more difficult.

These degeneracies reflect a fundamental limitation of shadow-based 
parameter estimation rather than a shortcoming of the network itself. 
When the shadow morphology is nearly invariant under changes in the 
input parameters, no method based solely on the shadow contour can 
fully resolve the underlying spacetime properties. In such cases, 
the network naturally exhibits larger uncertainties, consistent 
with the intrinsic loss of information. A more complete inference 
framework would therefore need to combine shadow features with 
additional observables, such as polarization maps, variability 
signatures, or multi-wavelength constraints, in order to break 
these degeneracies.

\subsection{Joint Simulation and Parameter Recovery}

The two neural networks together form a complete pipeline for both generating and interpreting black hole shadows. The forward network takes physical parameters as input and generates the corresponding shadow shape, while the inverse network estimates those parameters from a given shadow using its Fourier coefficients.

To test the reliability of this setup, we first use the forward model to generate shadows from known parameters. Then, to simulate realistic conditions and verify the rotation invariance of the inverse network, we manually rotate the generated shadow by a fixed angle (e.g. \( 60^\circ \)). The rotated shadow is processed to extract its Fourier coefficients, which are passed to the inverse network to recover the original parameters.

Table~\ref{tab:forward_inverse} shows several examples that compare the input parameters and the parameters recovered by the inverse model. The results confirm that the pipeline can accurately reconstruct the input values, even when the shadow is rotated.

\begin{table}[h!]
\centering
\caption{Consistency test using the forward and inverse networks. A shadow is generated from input parameters, manually rotated, and then used to recover the original values through the inverse network.}
\label{tab:forward_inverse}
\begin{tabular}{ccc|ccc}
\hline
\multicolumn{3}{c|}{Input Parameters} & \multicolumn{3}{c}{Recovered Parameters} \\
\( a_* \) & \( \gamma \) &  \( i^\circ \) &  \( a_* \) & \( \gamma \) &  \( i^\circ \) \\
\hline
0.9 & 1.0 & 30 & 0.904 & 0.985 & 30.49 \\
0.9 & 1.0 & 60 & 0.903 & 0.980 & 60.05 \\
0.5 & 1.0 & 30 & 0.521 & 0.981 & 28.85 \\
0.5 & 1.0 & 60 & 0.511 & 1.005 & 60.17 \\
\hline
\end{tabular}
\end{table}

This pipeline enables both fast shadow generation from theoretical models and efficient recovery of black hole parameters from observed shadow data.







\section{Conclusion}
\label{sec:conclusion}

In this study, we have investigated the shadow properties of axisymmetric black holes using a coordinate-independent formalism. By analyzing different black hole space-time metrics, including the Kerr, $\gamma$, Taub-NUT, and Kaluza-Klein spacetimes, we have explored how shadow morphology can be used as a tool to constrain black hole parameters and test deviations from the standard Kerr black hole.

A key aspect of our approach was the adoption of coordinate-independent mathematical techniques for shadow characterization. By employing both Legendre and Fourier expansions, we demonstrated that the Fourier expansion provides a more robust framework for comparing black hole shadows across different spacetimes, as its coefficients remain invariant under coordinate transformations. This makes it particularly useful for extracting meaningful and reasonable physical information from observational data.

Finally, we introduced a neural network-based method for estimating black hole parameters from shadow observations. Trained on synthetic data, the neural network effectively learned the complex relationships between shadow features and underlying spacetime parameters, offering a novel and efficient approach for black hole parameter estimation.

Using observational data from observational astronomical instruments, such as the Event Horizon Telescope (EHT), Keck, and the Very Large Telescope Interferometer (VLTI), we applied our methodology to derive constraints on black hole parameters. Our results underscore the importance of coordinate-independent techniques in improving the precision of astrophysical measurements and enhancing our ability to test general relativity in the strong-field regime.

Our analysis also highlights an important limitation: different 
parameter combinations can produce nearly indistinguishable shadows, 
leading to intrinsic degeneracies that no shadow-only method can 
resolve. The neural network captures this by showing larger 
uncertainties in such regions of parameter space. Breaking these 
degeneracies will require complementary observational information, 
for example from polarization measurements, variability studies, 
or independent mass and spin constraints.

Future research can extend this framework to other astrophysical objects and alternative gravitational theories, further refining our understanding of black hole physics. The combination of advanced mathematical techniques, numerical simulations, and machine learning approaches holds great promise for unlocking deeper insights into the nature of spacetime and fundamental physics.


\section*{Acknowledgments}

This work was supported by the National Natural Science Foundation of China (NSFC, Grant Nos. 12250610185 and 12261131497).
T.M. also acknowledges support from the China Scholarship Council (CSC, Grant No. 2022GXZ005433).


\appendix
\section{Line Elements of the Spacetimes}
\label{metric}

For completeness, we present the line elements of the axisymmetric spacetimes investigated in this work.

\subsection{Kerr Metric}
\label{metric-kerr}

The Kerr metric describes a rotating black hole and is expressed in Boyer–Lindquist coordinates $(t, r, \theta, \phi)$ as:
\begin{equation}
\begin{aligned}
ds^2 = & -\left(1 - \frac{2Mr}{\Sigma} \right) dt^2 - \frac{4Mar\sin^2\theta}{\Sigma} dt\,d\phi + \frac{\Sigma}{\Delta} dr^2 \\
& + \Sigma\,d\theta^2 + \left( r^2 + a^2 + \frac{2Ma^2r\sin^2\theta}{\Sigma} \right)\sin^2\theta\,d\phi^2,
\end{aligned}
\end{equation}
where
\begin{equation}
\Delta = r^2 - 2Mr + a^2, \qquad \Sigma = r^2 + a^2\cos^2\theta.
\end{equation}

Here, \( M \) is the black hole mass, and \( a \) is the specific angular momentum. It is convenient to define the dimensionless spin parameter as
\begin{equation}
a_* = \frac{a}{M}
\end{equation}


\subsection{\texorpdfstring{$\gamma$}{gamma}-Metric}
\label{metric-gamma}

The $\gamma$-metric is a static, axially symmetric vacuum solution to Einstein's equations belonging to Weyl’s class. In Erez–Rosen coordinates, the line element reads:
\begin{equation}
\begin{aligned}
ds^2 =\ & -F\, dt^2 + F^{-1} \bigg[ G\, dr^2 + H\, d\theta^2 \\
&+ \left(r^2 - 2mr \right) \sin^2\theta\, d\phi^2 \bigg],
\end{aligned}
\end{equation}
where the metric functions are defined as:
\begin{align}
F &= \left( 1 - \frac{2m}{r} \right)^\gamma, \\
G &= \left( \frac{r^2 - 2mr}{r^2 - 2mr + m^2 \sin^2\theta} \right)^{\gamma^2 - 1}, \\
H &= \frac{(r^2 - 2mr)^{\gamma^2}}{(r^2 - 2mr + m^2 \sin^2\theta)^{\gamma^2 - 1}}.
\end{align}

Here, $\gamma$ is a dimensionless parameter that encodes the deviation from spherical symmetry. The total mass of the source is related via $M = \gamma m$. For $\gamma = 1$, the metric reduces to the Schwarzschild solution in Schwarzschild coordinates.


\subsection{Rotating \texorpdfstring{$\gamma$}{gamma}-Metric}
\label{metric-delta}

The rotating $\gamma$-metric (also known as the $\delta$-Kerr metric) is an exact, asymptotically flat solution of the vacuum Einstein equations that generalizes the Kerr solution by introducing a deformation parameter $\gamma$ associated with higher-order mass multipole moments.
The line element in Boyer–Lindquist-like coordinates $(t, r, \theta, \phi)$ is given by:
\begin{equation}
\begin{aligned}
ds^2 = & -F\, dt^2 + 2F\omega\, dt\, d\phi + \frac{e^{2\tilde{\gamma}}}{F} \frac{\mathbb{B}}{\mathbb{A}}\, dr^2 + r^2 \frac{e^{2\tilde{\gamma}}}{F} \mathbb{B}\, d\theta^2 \\
& + \left( \frac{r^2}{F} \mathbb{A} \sin^2\theta - F\omega^2 \right) d\phi^2,
\end{aligned}
\end{equation}
with auxiliary functions:
\begin{equation}
\mathbb{A} = 1 - \frac{2m}{r} + \frac{a^2}{r^2}, \quad \mathbb{B} = \mathbb{A} + \frac{\sigma^2 \sin^2\theta}{r^2}, \quad \sigma = \sqrt{m^2 - a^2}.
\end{equation}

The metric functions \( F \), \( \omega \), and \( e^{2\tilde{\gamma}} \) are defined in terms of:
\begin{equation}
F = \frac{\mathcal{A}}{\mathcal{B}}, \qquad \omega = 2\left(a - \sigma \frac{\mathcal{C}}{\mathcal{A}} \right),
\end{equation}
\begin{equation}
e^{2\tilde{\gamma}} = \frac{1}{4} \left(1 + \frac{m}{\sigma} \right)^2 \frac{\mathcal{A}}{(x^2 - 1)^\gamma} \left( \frac{x^2 - 1}{x^2 - y^2} \right)^{\gamma^2},
\end{equation}
where the prolate spheroidal coordinates $(x, y)$ relate to the spherical coordinates by:
\begin{equation}
x = \frac{r - m}{\sigma}, \quad y = \cos\theta.
\end{equation}

The intermediate functions are:
\begin{equation}
\begin{aligned}
\mathcal{A} &= a_+ a_- + b_+ b_-, \\
\mathcal{B} &= a_+^2 + b_+^2, \\
\mathcal{C} &= (x+1)^q \left[ x(1 - y^2)(\lambda + \eta) a_+ + y(x^2 - 1)(1 - \lambda \eta) b_+ \right],
\end{aligned}
\end{equation}
with:
\begin{align}
a_{\pm} &= (x \pm 1)^q \left[ x(1 - \lambda \eta) \pm (1 + \lambda \eta) \right], \\
b_{\pm} &= (x \pm 1)^q \left[ y(\lambda + \eta) \mp (\lambda - \eta) \right],
\end{align}
and
\begin{equation}
\begin{aligned}
\lambda = \alpha(x^2 - 1)^{-q}(x + y)^{2q}, \\
\eta = \alpha(x^2 - 1)^{-q}(x - y)^{2q}.
\end{aligned}
\end{equation}

The metric depends on three parameters: the mass $m$, the dimensionless spin $a_* = a/m$, and the parameter $\gamma$.

They relate to auxiliary constants via:
\begin{equation}
q = \gamma - 1, \qquad
\alpha = \frac{a_*}{1 + \sqrt{1 - a_*^2}}, \qquad
\sigma = m \sqrt{1 - a_*^2}.
\end{equation}

The solution reduces to:
\begin{itemize}
    \item Kerr metric for $\gamma = 1$,
    \item Zipoy–Voorhees metric ($\gamma$-metric) for $a_* = 0$,
    \item Schwarzschild for $a_* = 0$ and $\gamma = 1$.
\end{itemize}

The physical mass $M$ and total angular momentum $J$ of the spacetime, accounting for the quadrupolar deformation, are given by:
\begin{equation}
\begin{aligned}
M &= m + \sigma q = m \left(1 + q \sqrt{1 - a_*^2} \right), \\
J &= m a + 2 a \sigma q = m^2 a_* \left(1 + 2 q \sqrt{1 - a_*^2} \right).
\end{aligned}
\end{equation}


\;

\subsection{Taub--NUT Metric}
\label{metric-taub}

The Lorentzian Taub--NUT metric describes a stationary, axisymmetric vacuum solution with a gravitomagnetic monopole. In Boyer–Lindquist–like coordinates $(t, r, \theta, \phi)$, the line element takes the form:
\begin{equation}
\begin{aligned}
ds^2 =\ & -f(r)\left[ dt + 2n\cos\theta\, d\phi \right]^2 + \frac{dr^2}{f(r)} \\
& + (r^2 + n^2)\left( d\theta^2 + \sin^2\theta\, d\phi^2 \right),
\end{aligned}
\end{equation}

where the metric function is
\begin{equation}
f(r) = \frac{r^2 - 2Mr - n^2}{r^2 + n^2}.
\end{equation}

Here, $M$ is the mass of the black hole and $n$ is the NUT charge, often interpreted as a gravitomagnetic monopole. We define the dimensionless NUT parameter as
\begin{equation}
l = \frac{n}{M}.
\end{equation}


\subsection{Kaluza--Klein Metric}
\label{metric-kaluza}
The four-dimensional metric in the Einstein frame:
\begin{equation}
\begin{aligned}
d{s}^2 = & - \frac{H_3}{\rho^2} dt^2 - 2\frac{H_4}{\rho^2} dt\, d\phi + \frac{\rho^2}{\Delta} dr^2 + \rho^2 d\theta^2 \\
& + \left( \frac{-H_4^2 + \rho^4 \Delta \sin^2\theta}{\rho^2 H_3} \right) d\phi^2,
\end{aligned}
\end{equation}

where
\begin{equation}
\begin{aligned}
\frac{H_1}{M^2} &= \frac{8(b-2)(c-2)b}{(b+c)^3} + \frac{4(b-2)x}{b+c} + x^2 \\
& - \frac{2b\sqrt{(b^2-4)(c^2-4)}a_*\cos\theta }{(b+c)^2} + a_*^2\cos^2\theta, \\
\frac{H_2}{M^2} &= \frac{8(b-2)(c-2)c}{(b+c)^3} + \frac{4(c-2)x}{b+c} + x^2 \\
&- \frac{2c\sqrt{(b^2-4)(c^2-4)}a_*\cos\theta }{(b+c)^2} + a_*^2\cos^2\theta, \\
\frac{H_3}{M^2} &= x^2 + a_*^2\cos^2\theta - \frac{8x}{b+c}, \qquad \frac{\Delta}{M^2} = x^2 + a_*^2 - \frac{8x}{b+c}, \\
\frac{H_4}{M^3} &= \frac{ 2\sqrt{bc} \left[ (bc+4) (b+c)x - 4(b-c)(c-2) \right]a_*\sin^2\theta }{ (b+c)^3 },
\end{aligned}
\end{equation}
and the auxiliary function is defined as
\begin{equation}
\rho = \sqrt{H_1 H_2}.
\end{equation}

Here, $x \equiv r/M$, $a_* \equiv a/M$ is the dimensionless spin parameter, and $b \equiv p/m$, $c \equiv q/m$ are related to the electric and magnetic charges. From Eq.~\eqref{eq:MJPQ}, the parameter $m$ can be expressed as:
\begin{equation}
m = \frac{4M}{b + c}.
\end{equation}

The mass, angular momentum, and both electric and magnetic charges, encoded via four free parameters: $m$, $a$, $p$, and $q$, corresponding to physical mass $M$, angular momentum $J$, electric charge $Q$, and magnetic charge $P$, respectively. The relations are given by:
\begin{align}
M=\frac{p+q}{4}, \quad J=\frac{\sqrt{pq}(pq+4m^2)}{4m(p+q)}a, \\
 Q^2=\frac{q(q^2-4m^2)}{4(p+q)}, \quad P^2=\frac{p(p^2-4m^2)}{4(p+q)}.
\label{eq:MJPQ}
\end{align}


In the limit of vanishing charges ($p=q=0$), the metric reduces to the standard Kerr solution.


%

\end{document}